\documentclass{article}

\usepackage[final]{neurips_2023}

\usepackage[utf8]{inputenc} 
\usepackage[T1]{fontenc}    
\usepackage{hyperref}       
\usepackage{url}            
\usepackage{booktabs}       
\usepackage{amsfonts}       
\usepackage{nicefrac}       
\usepackage{microtype}      
\usepackage{xcolor}         
\usepackage{siunitx}
\usepackage{hyperref}
\usepackage{enumitem}
\usepackage{amsmath,amsfonts}
\usepackage{graphicx}
\usepackage{tcolorbox}

\title{AceMap: Knowledge Discovery through \\Academic Graph}

\author{
    Xinbing Wang$^{1,2}$~\thanks{Xinbing Wang is the corresponding author (\href{mailto:xwang8@sjtu.edu.cn}{xwang8@sjtu.edu.cn}).}, Luoyi Fu$^{2}$, Xiaoying Gan$^{1}$, Ying Wen$^{3}$, Guanjie Zheng$^{3}$, Jiaxin Ding$^{3}$, \\
    \textbf{Liyao Xiang$^{3}$, Nanyang Ye$^{3}$, Meng Jin$^{3}$, Shiyu Liang$^{3}$, Bin Lu$^{1}$, Haiwen Wang$^{2}$, Yi Xu$^{2}$,}\\
    \textbf{Cheng Deng$^{2}$, Shao Zhang$^{1}$, Huquan Kang$^{2}$, Xingli Wang$^{2}$, Qi Li$^{1}$, Zhixin Guo$^{1}$, Jiexing Qi$^{1}$,}\\
    \textbf{Pan Liu$^{1}$, Yuyang Ren$^{2}$, Lyuwen Wu$^{2}$, Jungang Yang$^{2}$, Jianping Zhou$^{1}$, Chenghu Zhou$^{3,4}$}\\
    $^1$ Department of Electronic Engineering, Shanghai Jiao Tong University\\
    $^2$ Department of Computer Science, Shanghai Jiao Tong University\\
    $^3$ John Hopcroft Center for Computer Science, Shanghai Jiao Tong University\\
    $^4$ Institute of Geographic Sciences and Natural Resources Research, Chinese Academy of Sciences\\
    \texttt{xwang8@sjtu.edu.cn}
}

\begin{document}
\begin{sloppypar}
\maketitle

\begin{abstract}
The exponential growth of scientific literature requires effective management and extraction of valuable insights. While existing scientific search engines excel at delivering search results based on relational databases, they often neglect the analysis of collaborations between scientific entities and the evolution of ideas, as well as the in-depth analysis of content within scientific publications. The representation of heterogeneous graphs and the effective measurement, analysis, and mining of such graphs pose significant challenges. To address these challenges, we present AceMap, an academic system designed for knowledge discovery from a graph perspective. We present advanced database construction techniques to build the comprehensive AceMap database with large-scale academic entities that contain rich visual, textual, and numerical information. AceMap also employs innovative visualization, quantification, and analysis methods to explore associations and logical relationships among academic entities. AceMap introduces large-scale academic network visualization techniques centered on nebular graphs, providing a comprehensive view of academic networks from multiple perspectives. In addition, AceMap proposes a unified metric based on structural entropy to quantitatively measure the knowledge content of different academic entities. Moreover, AceMap provides advanced analysis capabilities, including tracing the evolution of academic ideas through citation relationships and concept co-occurrence, and generating concise summaries informed by this evolutionary process. In addition, AceMap uses machine reading methods to generate potential new ideas at the intersection of different fields. Exploring the integration of large language models and knowledge graphs is a promising direction for future research in idea evolution. Please visit \url{https://www.acemap.info} for further exploration.

\end{abstract}

\keywords{Academic Graph \and Scientific Literature \and Knowledge Graph \and Academic Big Data}


\section{Introduction}

\begin{figure}
    \centering
    \includegraphics[width=\linewidth]{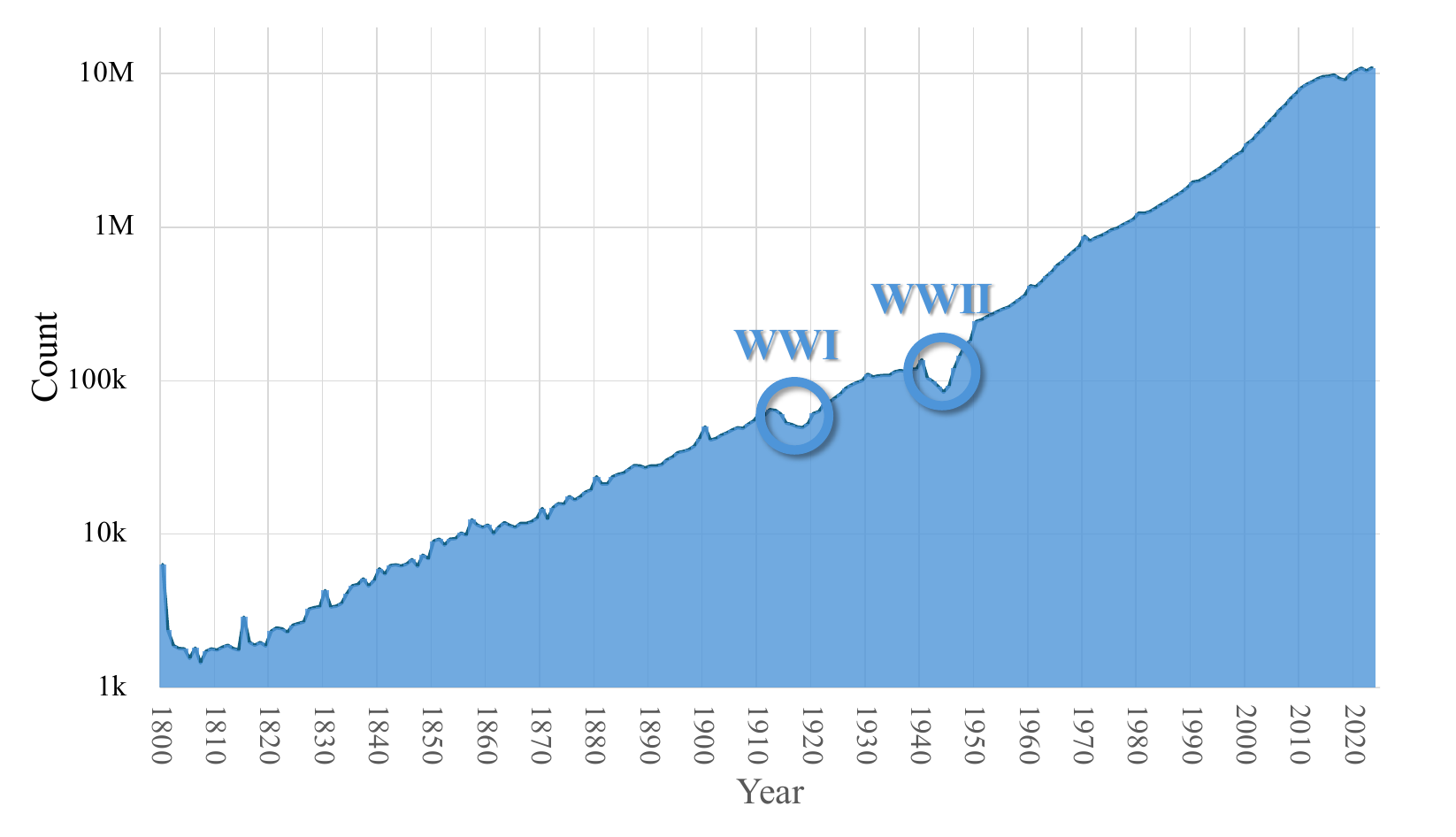}
    \caption{Annual Publication Trend in AceMap Database (1800-2023)}
    \label{fig:acemap_paper_count}
\end{figure}

The continuous advancement of scientific progress has triggered an exponential growth of scientific literature, fueling the expansion of knowledge on an unprecedented scale \cite{price1965networks,bornmann2015growth,hu2020exponential,10.1145/3543873.3587305}.
According to the National Science Board, the number of articles published in 2020 will be 2.9 million \cite{nsb2021}.
Our AceMap database provides insight into annual publication trends, as shown in Figure \ref{fig:acemap_paper_count}.
The graph shows the exponential growth in the number of articles published per year, with two exceptions around 1918 and 1944, coinciding with the periods of World Wars I and II. 
In particular, since 2020, the number of newly published articles has consistently exceeded ten million per year, indicating a significant increase in scholarly output.
Moreover, the popularization of pre-print platforms, such as arXiv \cite{ginsparg2011arxiv} and bioRxiv \cite{sever2019biorxiv}, and the widespread adoption of Open Access (OA) journals \cite{evans2009open} further accelerate the dissemination of knowledge.
The number of papers is growing rapidly, and it is infeasible and unproductive to read all of the countless papers. 
Researchers face a critical challenge when trying to discover valuable knowledge contained in large papers.
There is an urgent need for advanced tools and technologies to access, analyze, and exploit this vast corpus of information.

Currently, researchers rely primarily on academic search engines such as Google Scholar \cite{kousha2007google} and DBLP \cite{ley2009dblp} to conduct their research. While these platforms are useful for searching academic literature, they fail to capture the various interactions between different academic entities.  
As Steve Jobs once said: ``Creativity is just connecting things.'' 
In the academic world, collaboration and citation are essential to advancing scientific knowledge. They create complex and dynamic networks of papers, authors, institutions, and concepts that reflect the patterns and trends of scientific development. These networks provide deeper insights into the evolution and structure of knowledge than can be obtained from the relational model alone. Therefore, analyzing data from the literature using an academic graph is beneficial for gaining a better understanding of the science of science.

However, managing and analyzing literature data from a graph perspective presents challenges that require addressing four key questions: How can we effectively represent the complex and heterogeneous scholarly networks that result from the interactions among different scholarly entities? How can we represent the structure of large-scale scholarly networks in an intuitive way? How can we measure and compare the amount of knowledge held by different academic entities (e.g. papers, authors, institutions) based on their network properties? How can we extract and analyze knowledge from academic networks and present it in human-readable natural language? To address these challenges, we design AceMap, a platform for knowledge discovery from large scientific literature using academic graphs. 
In this paper, we summarize our contribution and experience in investigating AceMap, which includes four pieces of work as follows.

\begin{itemize}[leftmargin=0.5cm]
\item \textbf{Construction}: We developed a domain-wide academic literature knowledge graph database and platform on a \num{e8} scale that provides a comprehensive and streamlined approach to searching, ranking, and summarizing a variety of academic entities, including papers, authors, institutions, and concepts  (Section \ref{sec:construction}).
\item \textbf{Visualization}: We proposed a large-scale academic network visualization technique, called VSAN, which offers a comprehensive overview of the academic context and reveals the patterns and trends of knowledge development (Section \ref{sec:visualization}).
\item \textbf{Quantification}: We proposed KQI to measure the knowledge quantity of literature based on graph structural entropy, which enables a unified measurement of heterogeneous academic entities. These novel metrics help rank and compare academic entities based on their knowledge quantity (Section \ref{sec:quantification}).
\item \textbf{Analysis}: To track the evolution of academic ideas, we implemented text summarization techniques based on citation relationships and concept co-occurrence. We also used machine reading methods to generate cross-disciplinary innovative text using natural language processing techniques (Section \ref{sec:analysis}).
\end{itemize}

We have been working on AceMap since 2013, and have been dedicated to updating and refining it ever since. This report serves as a summary of our work over the past decade. The rest of the paper is organized as follows. In Section \ref{sec:overview}, we introduce the overview of AceMap, including the ontology design and the user interface. Section \ref{sec:construction} explains how we build AceMap database, i.e. AceKG, as a scientific knowledge graph. In Section \ref{sec:visualization}, we delve into the visualization of large-scale knowledge networks. In Section \ref{sec:quantification}, we present the quantification of different entities in heterogeneous academic networks from a graphical perspective. Finally, in Section \ref{sec:analysis}, we discuss how to express various types of link relationships, e.g. the evolution of academic concepts, in a human-readable natural language.

\section{Overview of AceMap}
\label{sec:overview}

\subsection{AceMap Ontology}

\begin{figure}
    \centering
    \includegraphics[width=\linewidth]{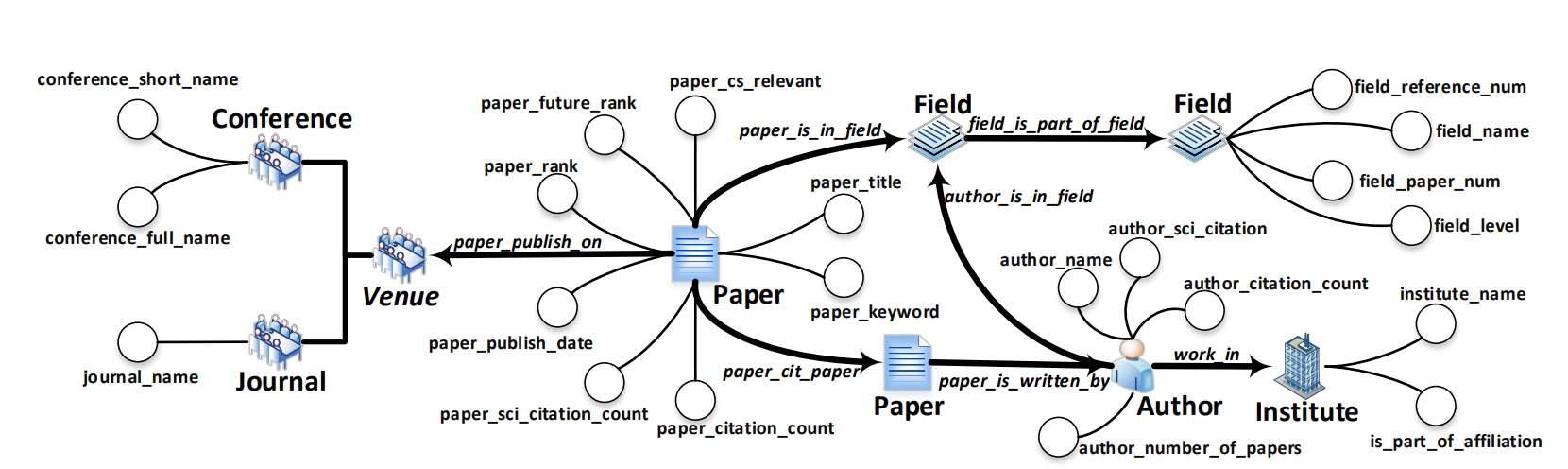}
    \caption{An overview of AceKG Ontology.}
    \label{fig:overview_ontology}
\end{figure}

The overview of the AceKG ontology is shown in figure \ref{fig:overview_ontology}.All objects (e.g. papers, institutes, authors) are represented as entities in AceKG. Two entities can have a relationship. Commonly used attributes of each entity, including numbers, dates, strings, and other literals, are also
literals are also represented.  Similar entities are grouped into classes. In total, AceKG defines 5 classes of academic entities: \textit{papers, authors, venues, fields of study and institutes}. The facts, including the frequently used properties of each entity and the relationships between entities, are described as triples in the knowledge graph; see the Table \ref{table:acekg-schema} for more details.

\begin{table}[htbp]
\caption{Ontology Design of AceMap Knowledge Graph}
\label{table:acekg-schema}
\centering
\begin{tabular}{llll}
\hline
\textbf{Class} & \textbf{Subject}    & \textbf{Predicate}            & \textbf{Object}     \\ \hline
Paper          & ace:Paper\_ID       & rdf:type                      & ace:Paper           \\
               & ace:Paper\_ID       & ace:paper\_citation\_count    & xsd:int             \\
               & ace:Paper\_ID       & ace:paper\_cs\_relevant       & xsd:boolean         \\
               & ace:Paper\_ID       & ace:paper\_future\_rank       & xsd:int             \\
               & ace:Paper\_ID       & ace:paper\_keyword            & xsd:string          \\
               & ace:Paper\_ID       & ace:paper\_rank               & xsd:int             \\
               & ace:Paper\_ID       & ace:paper\_publish\_date      & xsd:date            \\
               & ace:Paper\_ID       & ace:paper\_sci\_citation      & xsd:int             \\
               & ace:Paper\_ID       & ace:paper\_title              & xsd:string          \\
               & ace:Paper\_ID       & ace:paper\_is\_written\_by    & ace:Author\_ID      \\
               & ace:Paper\_ID       & ace:paper\_is\_in\_field      & ace:Field\_ID       \\
               & ace:Paper\_ID       & ace:paper\_publish\_on        & ace:Venue\_ID       \\ 
               & ace:Paper\_ID       & ace:paper\_cit\_paper         & ace:Paper\_ID       \\ \hline
Author         & ace:Author\_ID      & rdf:type                      & ace:Author           \\
               & ace:Author\_ID      & ace:author\_name              & xsd:string          \\
               & ace:Author\_ID      & ace:author\_citation\_count   & xsd:int             \\
               & ace:Author\_ID      & ace:author\_number\_of\_paper & xsd:int             \\
               & ace:Author\_ID      & ace:author\_sci\_citation     & xsd:int             \\
               & ace:Author\_ID      & ace:author\_is\_in\_field     & ace:Field\_ID       \\
               & ace:Author\_ID      & ace:work\_in                  & ace:Institute\_ID   \\ \hline
Institute      & ace:Institute\_ID   & rdf:type                      & ace:Institute       \\
               & ace:Institute\_ID   & ace:institute\_name           & xsd:string          \\ \hline
Venue          & ace:Venue\_ID       & rdf:type                      & ace:Conference      \\
               & ace:Venue\_ID       & ace:conference\_full\_name    & xsd:string          \\
               & ace:Venue\_ID       & ace:conference\_short\_name   & xsd:string          \\
               & ace:Venue\_ID       & rdf:type                      & ace:Journal         \\
               & ace:Venue\_ID       & ace:journal\_name             & xsd:string          \\ \hline
Field          & ace:Field\_ID       & rdf:type                      & ace:Field           \\
               & ace:Field\_ID       & ace:field\_name               & xsd:string          \\
               & ace:Field\_ID       & ace:field\_level              & xsd:string          \\
               & ace:Field\_ID       & ace:field\_papers\_num        & xsd:int             \\
               & ace:Field\_ID       & ace:field\_reference\_count   & xsd:int             \\
               & ace:Field\_ID       & ace:field\_is\_part\_of       & ace:Field\_ID       \\ \hline
\end{tabular}
\end{table}

\begin{description}[leftmargin=2em]
    \item[Entities.] Each entity has a class type, a highly abstracted entity. The entities  designed can be listed as follows:
    \begin{description}[leftmargin=0.8em]
        \item[Paper (ace:Paper)]  Representation of the academic papers. Entity \textit{ace:paper} has \textbf{8} data properties including title, key word, citation count and SCI citation count, rank and future rank, cs relevant as well as the publish date. 
    
        \item[Author (ace:Author)] Representation of the scholars. Entity \textit{author} has \textbf{4} data properties including author's name, number of papers, SCI citation and count.
    
        \item[Institute (ace:Institute)] Representation of the institutes. Entity \textit{institute} has \textbf{1} data properties, a name.
    
        \item[Venue (ace:Venue)] Representation of the conferences and journals. Entity \textit{venue} has \textbf{2} categories including \textit{conference} and  \textit{journal}, which contains the attributes the short name and full name of conference, journal name, respectively.
    
        \item[Field (ace:Field)] Representation of the fields in which the author works. Entity \textit{field} has \textbf{4} data properties including field's name, level, number of reference and number of paper.\\
    \end{description}

    \item[Relations.] Also can be deemed for entities' object properties. The axioms corresponding to relations are defined as following:
    \begin{description}[leftmargin=0.8em]
        \item[ace:paper\_cite\_paper] connects two paper entities, which means the former paper refers to the latter.
    
        \item[ace:paper\_is\_written\_by] connects entity paper and entity author, which illustrates who wrote the paper.
    
        \item[ace:work\_in] connects entity author and entity institute, and it shows that which institute the author work in.
    
        \item[ace:paper\_is\_in\_field] connects entity paper and entity field, showing that which field the paper was published.
    
        \item[ace:author\_is\_in\_field] connects entity author and entity field, showing that which field the author belongs to.
    
        \item[ace:paper\_publish\_on] connects entity paper and entity venue, which means the venue on which the paper was published.
    
        \item[ace:field\_is\_part\_of\_field] connects two field entities. It means that the former domain is part of the latter.
    \end{description}
\end{description}

AceMap collects data from multiple disciplines, including more than 220 million papers metadata with 103 million authors, 767 thousand research fields, 19 thousand academic institutes, 50 thousand journals, 4 thousand conferences, and 113 thousand affiliations. 
The AceMap Knowledge Graph (AceKG) provides comprehensive descriptions of 114.30 million academic entities, following a standardized ontology. In total, AceKG comprises a repository of 3.13 billion pieces of relationship information.

\subsection{User Interface of AceMap}

\begin{figure}[ht]
    \centering
    \includegraphics[width=\linewidth]{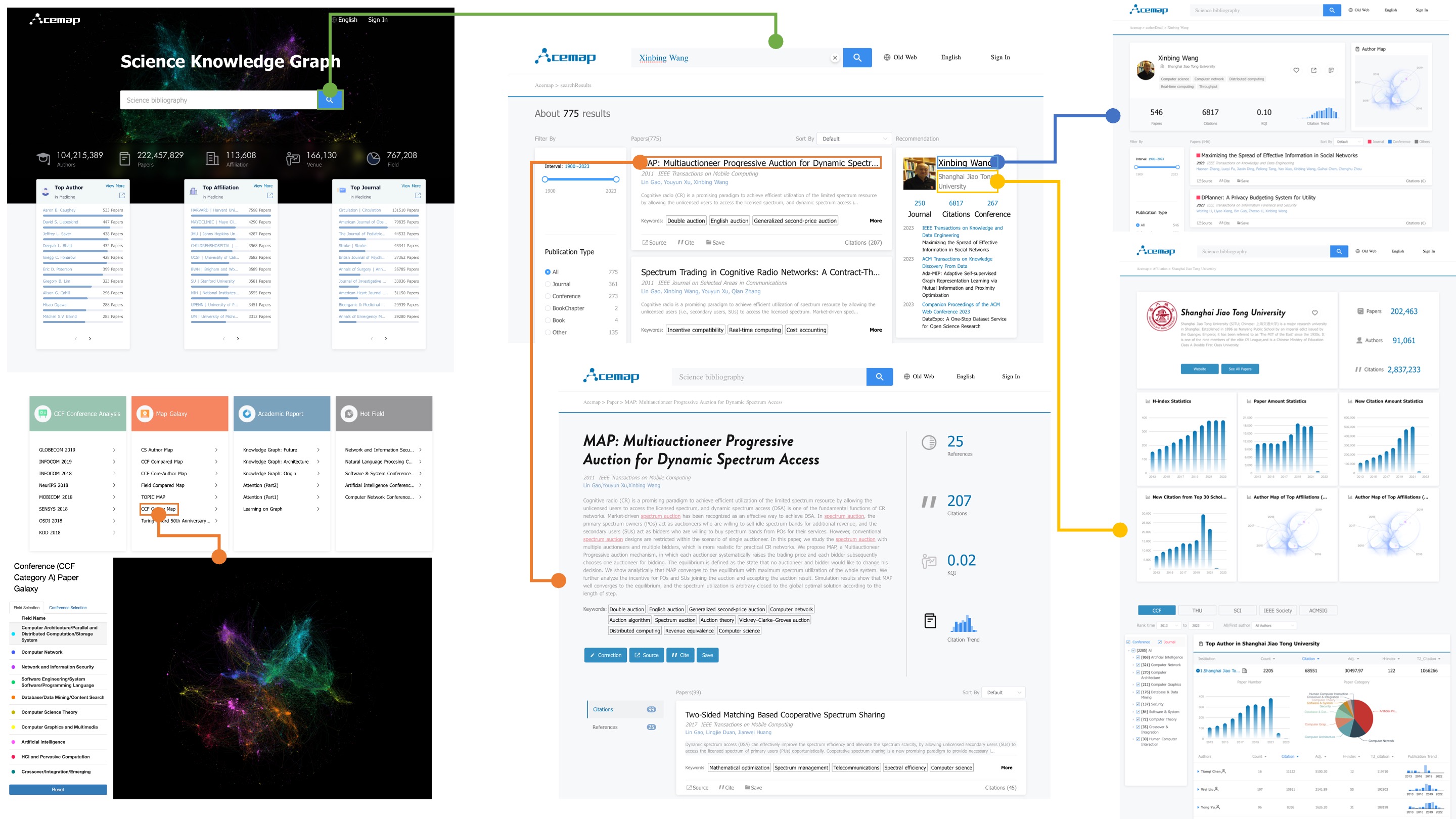}
    \caption{The user interface of AceMap for academic search and visualization.}
    \label{fig:user_interface_acemap}
\end{figure}

Figure \ref{fig:user_interface_acemap} shows the user interface of AceMap. At the top of the home page is a search box that allows users to search for papers, authors, institutions, and more. In addition, the homepage offers a number of useful tools. For example, users can click on certain links to explore galaxy-like visualizations of academic graphs. By clicking on the search icon, users are presented with paper results in the center and recommended content on the right. By clicking on the appropriate links, users can conveniently access detailed information about the papers, authors, institutions, etc.

\begin{figure}[h]
    \centering
    \includegraphics[width=1.0\linewidth]{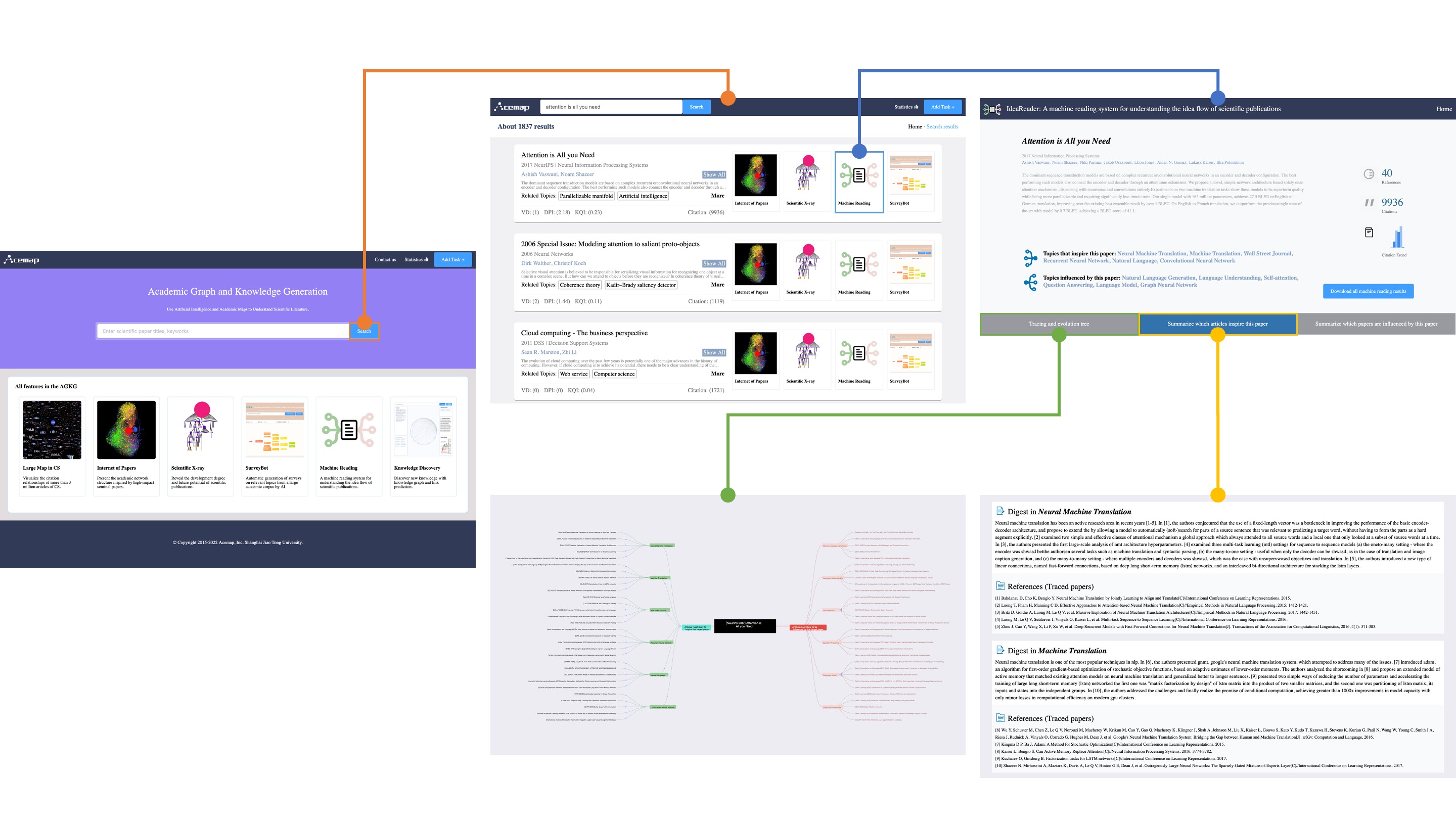}
    \caption{The user interface of AceMap for exploring the idea flow in academic papers.}
    \label{fig:user_interface_ideareader}
\end{figure}

The interface shown in Figure 
\ref{fig:user_interface_ideareader} allows users to explore the inheritance relationships of academic ideas within a paper. A search box is conveniently located at the top of the interface. Users can enter the title of a paper in the search box and click the search icon to view the search results. By selecting the `Machine Reading' option of the target search result, users can access the IdeaReader page. This page has three tabs available, each offering distinct views. The first tab provides a visual representation of the tracing tree and the evolution tree, allowing users to trace the origins and influence of the paper. The second and third tabs offer concise textual summaries of the tracing tree and the evolution tree, respectively. These summaries provide users with a quick overview of the key information contained within the trees.

\begin{figure}[h]
    \centering
    \includegraphics[width=1.0\linewidth]{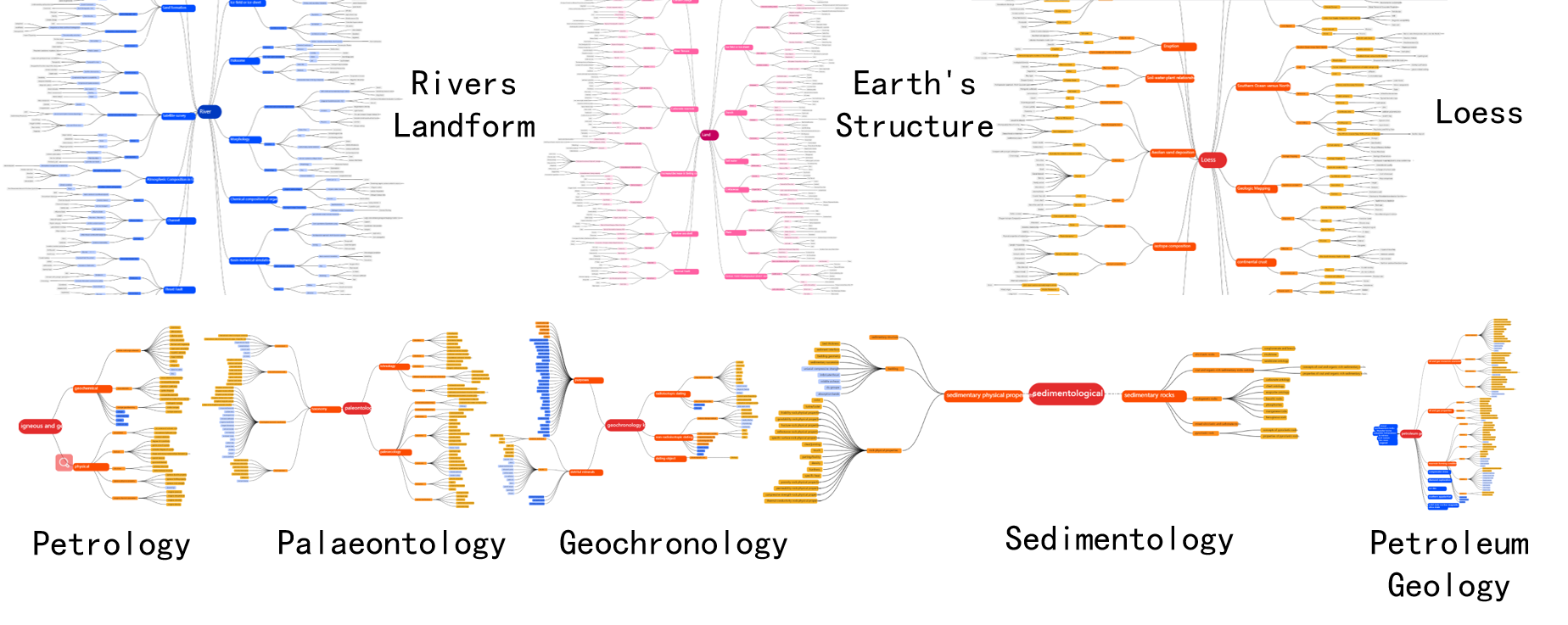}
    \caption{Several Topic Tree constructed by AceMap in the field of Geoscience.}
    \label{fig:topictree}
\end{figure}

\section{Construction: Multi-model scientific knowledge graph}
\subsection{Topic Tree Construction}
AceMap hierarchically represents the various concepts or entities existing in the field of geosciences and the hierarchical analogical relationships between them based on a topic model, which is shown in Figure~\ref{fig:topictree}.
The Topic Tree \cite{chen2017latent} aims to describe various concepts or entities existing in the real world and the hierarchical analogical relationships between them, and presents a tree-like structure overall, with nodes representing concepts or entities and edges consisting of relationships. Specifically, nodes represent knowledge concepts or entities. Concepts refer to the set of entities with the same characteristics or abstract knowledge terms, such as researchers, companies, organizations, and artificial intelligence, machine learning, knowledge graphs, etc. Relationships describe the hierarchical affiliation between nodes, including the juxtaposition and subordination (genus) relationships between entity concepts. For example, machine learning is a subset of artificial intelligence, and computer vision is a subset of machine learning. Relationships in a Topic Knowledge Tree are mathematically formalized as a function that maps $k$ nodes to a Boolean value.

Hierarchical hidden variables are used to model co-occurrence patterns of entities. The state of a hidden variable represents a collection of papers under a research topic, and the concatenated edges between hidden variables represent the subordination between topics. The process of constructing the topic tree is divided into three steps. Firstly, to ensure objectivity, we detect the co-occurrence relationship of entities in the literature and construct a co-occurrence network using the relationship of words in different theses. We then use a community detection algorithm to divide the word variables into several patterns. No changes in content have been made. This creates tight word relationships within the patterns and sparse relationships between them, allowing for the construction of leaf nodes of the topic tree. The language used is clear, concise, and value-neutral, with a formal register and precise word choice. The text is grammatically correct and follows conventional academic structure and formatting. The algorithm detects the co-occurrence relationship between patterns, which can be regarded as hyperpoints. The algorithm detects the co-occurrence relationship between patterns, which can be regarded as hyperpoints. The algorithm detects the co-occurrence relationship between patterns, which can be regarded as hyperpoints. It is then iteratively utilized to build a tree structure containing multi-level topics. Finally, the themes in the patterns are extracted. The model utilizes a bottom-up algorithm for theme extraction. It calculates the mutual information between each keyword and its corresponding schema, filters out the keyword set that can represent the theme from the sub-schema, and finally obtains the schema theme by using the keyword set to fuzzy match with the entity names in the existing database. It calculates the mutual information between each keyword and its corresponding schema, filters out the keyword set that can represent the theme from the sub-schema, and finally obtains the schema theme by using the keyword set to fuzzy match with the entity names in the existing database.

\label{sec:construction}
\subsection{Academic Entity Extraction and Relation Construction}
\begin{figure}[h]
    \centering
    \includegraphics[width=0.8\linewidth]{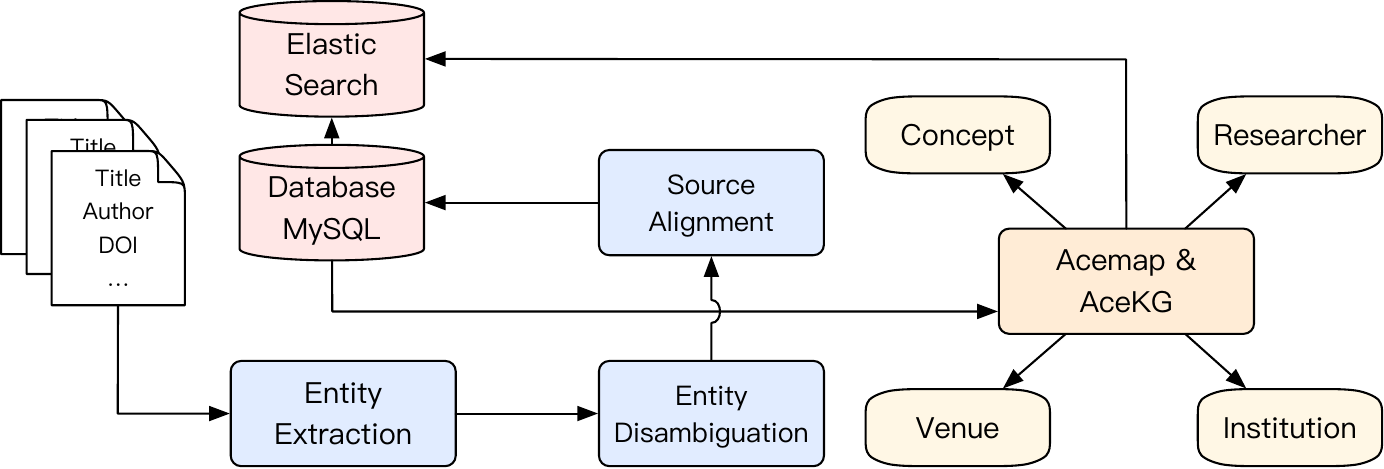}
    \caption{The pipeline of AceMap Construction}
    \label{fig:acemap_pipeline}
\end{figure}

As a novel platform to advance scientific research and discovery, AceMap relies on a foundation of various technical components to construct and maintain its scientific knowledge. In this part, we will delve into the key technical points involved in the construction of AceMap, including entity extraction, entity disambiguation, and source alignment. The pipeline is demonstrated in Figure~\ref{fig:acemap_pipeline}.

\paragraph{Entity Extraction} At the heart of AceMap's mission to advance scientific research and discovery is the fundamental process of concept entity extraction. This critical step involves identifying and extracting specific concept entities from the vast corpus of scientific literature. The entities cover a wide range from researchers and institutions to publications, topics and research areas. Using the extracted entities, AceMap is able to connect potential knowledge.

To accomplish this, AceMap uses a combination of natural language processing (NLP) techniques and machine learning algorithms. The process begins with data acquisition from a variety of sources, including academic journals, conference proceedings, preprint servers, and patent databases. This raw text data is then subjected to a series of preprocessing steps, such as tokenization, stemming, and stop word removal, to facilitate subsequent analysis. The core of the entity extraction is based on Named Entity Recognition (NER), a subset of NLP that focuses on identifying and classifying entities in text. AceMap uses state-of-the-art NER models trained on large labeled datasets spanning the scientific literature. In our initial version, these models use deep learning techniques, including bidirectional transformers such as BERT~\cite{devlin2018bert}, to achieve high accuracy in entity extraction. In addition, AceMap continuously refines its entity extraction process by leveraging domain-specific knowledge. This iterative approach ensures that the platform evolves to include new entities and adapts to changes in the scientific landscape over time.

\paragraph{Entity Disambiguation} In the scientific literature, entity disambiguation is a challenging but essential task. Given the inherent ambiguity of entity names and references, it is crucial to disambiguate entities in order to accurately map them to their real-world counterparts. AceMap addresses this challenge through a multi-step process.

First, the extracted entities are associated with unique identifiers whenever possible. For example, researchers can be linked to their ORCID iDs, institutions to their official identifiers, and publications to their DOIs. This linking helps reduce ambiguity by providing a standardized reference point. However, not all entities can be linked to unique identifiers, and many names are shared between different entities. To address this, AceMap uses disambiguation algorithms that take into account contextual information. These algorithms take into account factors such as co-authorship relationships, publication history, and research interests to probabilistically resolve entity references. To further improve classification accuracy, AceMap uses external knowledge sources such as author profiles, citation networks, and semantic similarity measures. By incorporating these contextual cues and external data, the platform can distinguish between entities with similar names, thereby reducing errors in entity mapping.

\paragraph{Source Alignment} Source alignment is the final piece of the puzzle in building AceMap KG database. Once entities have been extracted and disambiguated, the platform must align them across different data sources to create a coherent and interconnected knowledge graph.

This process involves identifying relationships and connections between entities across different sources. For example, linking a researcher to their affiliated institution, linking co-authors based on their collaboration history, and linking publications to their respective authors and institutions. To achieve this, AceMap uses a combination of graph-based algorithms, knowledge graph embeddings, and network analysis techniques. These methods help uncover latent relationships and patterns within the data, facilitating the alignment of entities in a way that reflects the true structure of the scientific ecosystem.

\subsection{Multi-model Data Extraction from Paper Content}

To better extract and manage data, we design and impletment an AI-in-the-loop system, DeepShovel \cite{zhang2022knowledgeshovel,zhang2022deepshovel,zhang2023geodeepshovel}, where researchers collaborate with AI in extracting data efficiently from literature to build a scientific knowledge base with high data quality without the dependence of data scientists 

\subsubsection{System Design}

The complete work of the system including the formative study, system description, and user study can be found at \cite{zhang2022knowledgeshovel,zhang2022deepshovel,zhang2023geodeepshovel}.
DeepShovel allows users to navigate through different tabs such as Meta, Text, Table, and Map to facilitate different aspects of data extraction. The process involves a sequential human-AI collaboration pipeline that enhances user interaction with the system.

To facilitate the extraction of meta information from articles, we have developed a module that autonomously retrieves details such as title, author(s), journal/conference, and additional meta information from the PDF file.
This feature helps users to preserve the data source within the database for future reference and citation in their research efforts.

Due to insufficient accuracy in fully automated extraction and processing of tables for scientific knowledge bases, manual handling of these tables becomes a tedious task. The system's human-AI collaboration pipeline enables rapid AI-assisted data acquisition, while relying on human judgment for final verification of data accuracy. This approach not only ensures maximum data accuracy, but also significantly reduces the number of steps compared to fully manual cell-by-cell data transcription.

In text extraction, weak supervision learning models and rules are used to perform named entity recognition tasks. This helps to highlight key terms in the text and attributes of the samples, helping users to incorporate these elements into their data sets. All pre-identified entities are prominently displayed, allowing users to easily find specific information.

We also offer a module that can recognize maps and determine the latitude and longitude for each point. DeepShovel allows users to adjust the detected latitude and longitude values and the placement of these coordinates on the map, further improving the accuracy of the results.

\subsubsection{System Impletmentation}

\paragraph{Meta Information Extraction} For each uploaded document, our system uses various parsing tools such as Grobid, Science Parse, and PdfFigures 2.0, each of which independently extracts meta information. The extracted data is then synthesized using a voting mechanism to ensure accuracy and completeness.

\paragraph{Table Extraction:} First, we use the object detection model Detectron2 \cite{wu2019detectron2}, trained on the TableBank dataset \cite{li2019tablebank}, a standard for table detection, to identify table regions. For each table, a set of rules is implemented to locate each cell. After user verification of the cell layout, Tesseract \cite{10.5555/1288165.1288167} is used to recognize text in each cell, culminating in the creation of a digitized table.

\begin{figure}
\centering
\includegraphics[scale=0.4]{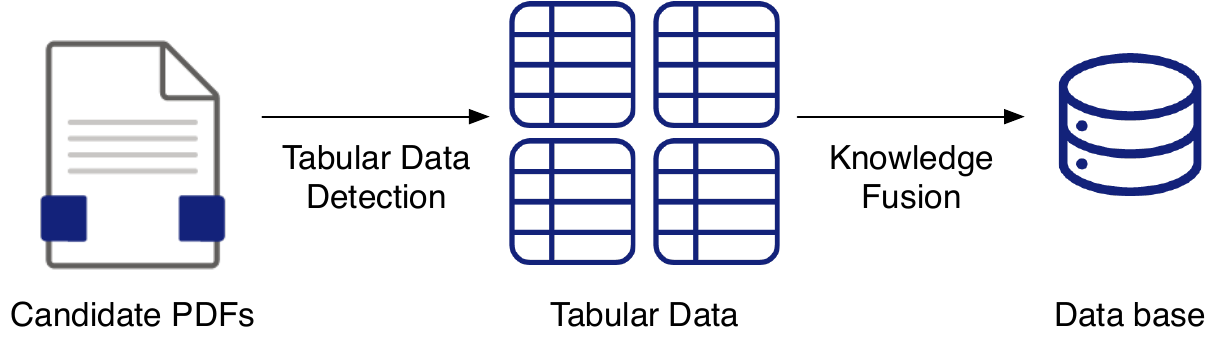}
\caption{An overview of tabular data collection workflow}
\label{workflow}
\end{figure}

\paragraph{Tabular Data Collection} Figure~\ref{workflow} presents an overview of the tabular data collection workflow, encompassing three primary components: (1) \textit{PDF Parsing}, (2) \textit{Tablular Data Detection}, (3) \textit{Knowledge Fusion}.
To enhance the efficiency of data retrieval, we implement a preprocessing step on all PDF files, as illustrated in Figure~\ref{workflow}. This step involves extracting and parsing pertinent metadata from each file. Subsequently, an index is generated for each PDF, grounded on the extracted metadata. This indexed metadata facilitates precise keyword searches, enabling us to efficiently isolate the relevant PDF data from an extensive collection of files.

\begin{figure}
\centering
\includegraphics[scale=0.36]{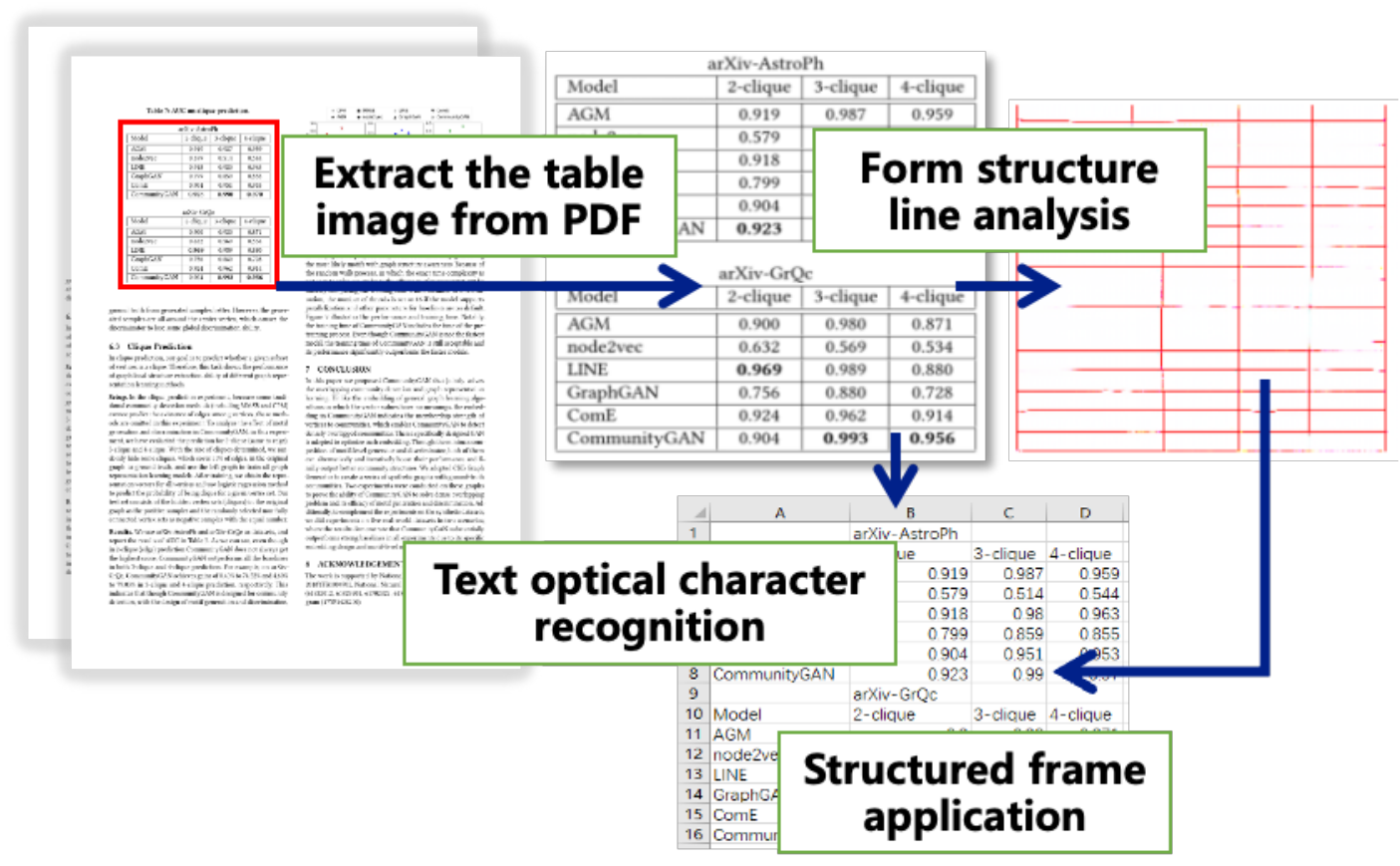}
\caption{An overview of tabular data detection}
\label{tableflow}
\end{figure}

After preprocessing the PDF files, we extract the visual elements and then convert them into XML files composed of structured tokens~\cite{guo2023geoimagecut, guo2023towards}. As shown in Figure~\ref{tableflow}, we use neural network-based techniques specifically tailored for tabular data recognition to extract target elements from these XML files. The challenge of tabular data recognition lies in its complexity, given the wide range of structures encountered. To improve the accuracy of table extraction, we fine-tuned the YOLOv3 model using the Tablebank dataset \cite{li2020tablebank}. 

After data extraction, we systematically collect and classify potential entity names and match them against a standardized lexicon to address name disambiguation according to the provided keyword list. To enhance the schema adaptation process, we use BERT (Bidirectional Encoder Representations from Transformers) \cite{devlin2018bert} to encode each entity name into a high-dimensional vector, thereby generating a comprehensive representation. In the knowledge fusion domain, the designated keywords are normalized by evaluating the similarity between the user's preference vector and the standardized entity names. As a result, the query results are ready for immediate automated download.

\paragraph{Text Extraction} To retrieve academic entities from PDF-formatted papers, we start by using PDFFigures 2.0 \cite{clark2016pdffigures} to separate text sections from the source files. Subsequently, a combination of defined rules and the natural language processing tool spaCy \cite{spacy2} is used for the automated extraction of various types of entities from these text sections.

\paragraph{Location Extraction in Figure} Users have the option to select any map region of interest. Our system then identifies longitude and latitude markings at the map's edges to ascertain the map's coordinate boundaries. When users select a point on the map, the precise coordinates of that location are automatically calculated and recorded.

\begin{figure}[!t]
\centering
\includegraphics[scale=0.42]{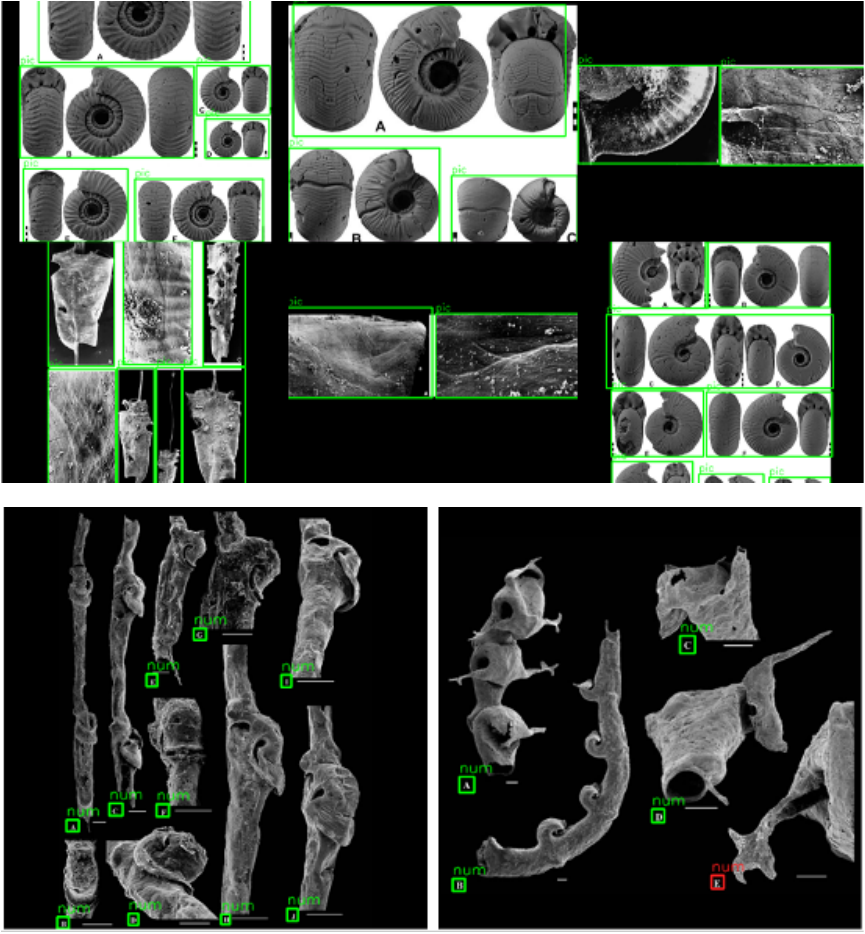}
\caption{An example of object detection~\cite{guo2023geoimagecut}.}
\label{objde}
\end{figure}

\paragraph{Image Cut}
In the sub-image detection effort, we solicited the expertise of domain specialists to annotate a dataset of 500 images. Subsequently, the dataset was fine-tuned using YOLO4 \cite{jiang2022review}, an advanced CNN-based object detection framework renowned for its accuracy and computational power. It is imperative to note that the acquisition of sub-image data requires the alignment of the sub-image label with its corresponding caption. To optimize the sub-image label identification procedure, we designed a network that synergistically combines a label object detection model with a text recognition model. The YOLO4 framework was used for label object detection, while the CRNN model \cite{fu2017crnn} was used for text recognition. As shown in Figure~\ref{objde}, the top part shows an example of sub-image recognition, and the bottom part shows an example of label recognition.

Using our technology, we have successfully harnessed a vast amount of information. 
According to our research, it is estimated that there are over 20,000,000 figures, 3,000,000 maps, 3,800,000 tables, and 12,000,000 formulas contained in papers in the field of geoscience alone.
In addition, AceMap includes not only geoscience literature, but also a vast collection of papers from various fields. 
These articles include a wealth of graphical, textual, and data information, providing a rich landscape for further exploration and discovery.

\section{Observation: Evolution of scientific literature}
\label{sec:visualization}

With the rapid growth of science and technology, the number of research papers has increased exponentially. Much like how the Internet of Things (IoT) is connecting the world in a more streamlined manner, one might wonder how the network formed by this vast sea of academic papers would appear. The challenge lies in the fact that most existing visualization methods \cite{ForceAtlas2, Fruchterman, Yifan} can only handle networks with node sizes up to hundreds of thousands. This limitation falls far short of the scale of academic networks, which often consist of millions or even more nodes. In response to this need, we are driven to overcome this scalability limitation and have developed a novel visualization method specifically tailored for super-large academic networks, which we refer to as VSAN (Visualizing Super-Large Academic Networks) \cite{li2024vsan}. The description of the VSAN algorithm pipeline is as follows (see Figure~\ref{fig:qili_f4}):

\begin{figure}[h]
    \centering\textbf{}
    \includegraphics[width=1.0\linewidth]{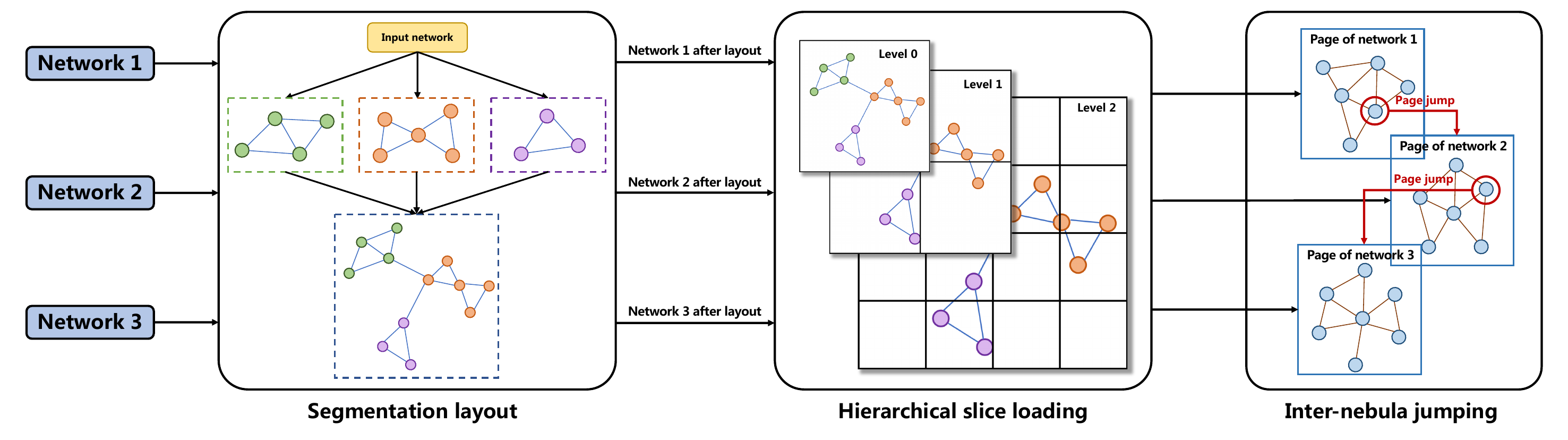}
    \caption{The pipeline of the VSAN algorithm}
    \label{fig:qili_f4}
\end{figure}

\paragraph{Graph segmentation}The initial step involves breaking down the graph based on the original data using a heuristic algorithm. This segmentation process employs a community partitioning algorithm with linear complexity, which avoids the need to directly apply traditional layout algorithms to a network with millions of nodes.

\paragraph{Generation of inter-block layout} Leveraging the results of the graph segmentation, we employ a force-directed model to generate the inter-block layout of subgraphs. This step aims to establish a clear macro structure for the final visualization outcome. Since the network size has been significantly reduced during the equivalence process, it has no adverse impact on the overall algorithm's efficiency.

\paragraph{Generation of subgraph layout}Subsequently, the force-directed model is used to lay out the subgraphs individually, yielding optimal solutions for subgraph layout. The segmented subgraph's size is substantially smaller than the original, enabling traditional layout algorithms to produce clear results and ensuring the accurate depiction of local microstructures.

\paragraph{Stitching of subgraphs}The subgraphs are stitched together based on their inter-block layouts to form the preliminary graph. This step primarily involves translating the positions of nodes in each subgraph without complex calculations, preserving the algorithm's efficiency.

\paragraph{Fine-tuning}To achieve the optimal layout for the entire graph, we fine-tune the preliminary graph's layout using the force-directed model. This process entails only a small number of iterations on the current layout and is sometimes optional without a significant impact on overall algorithm efficiency.

\begin{figure}[h]
    \centering\textbf{}
    \includegraphics[width=1.0\linewidth]{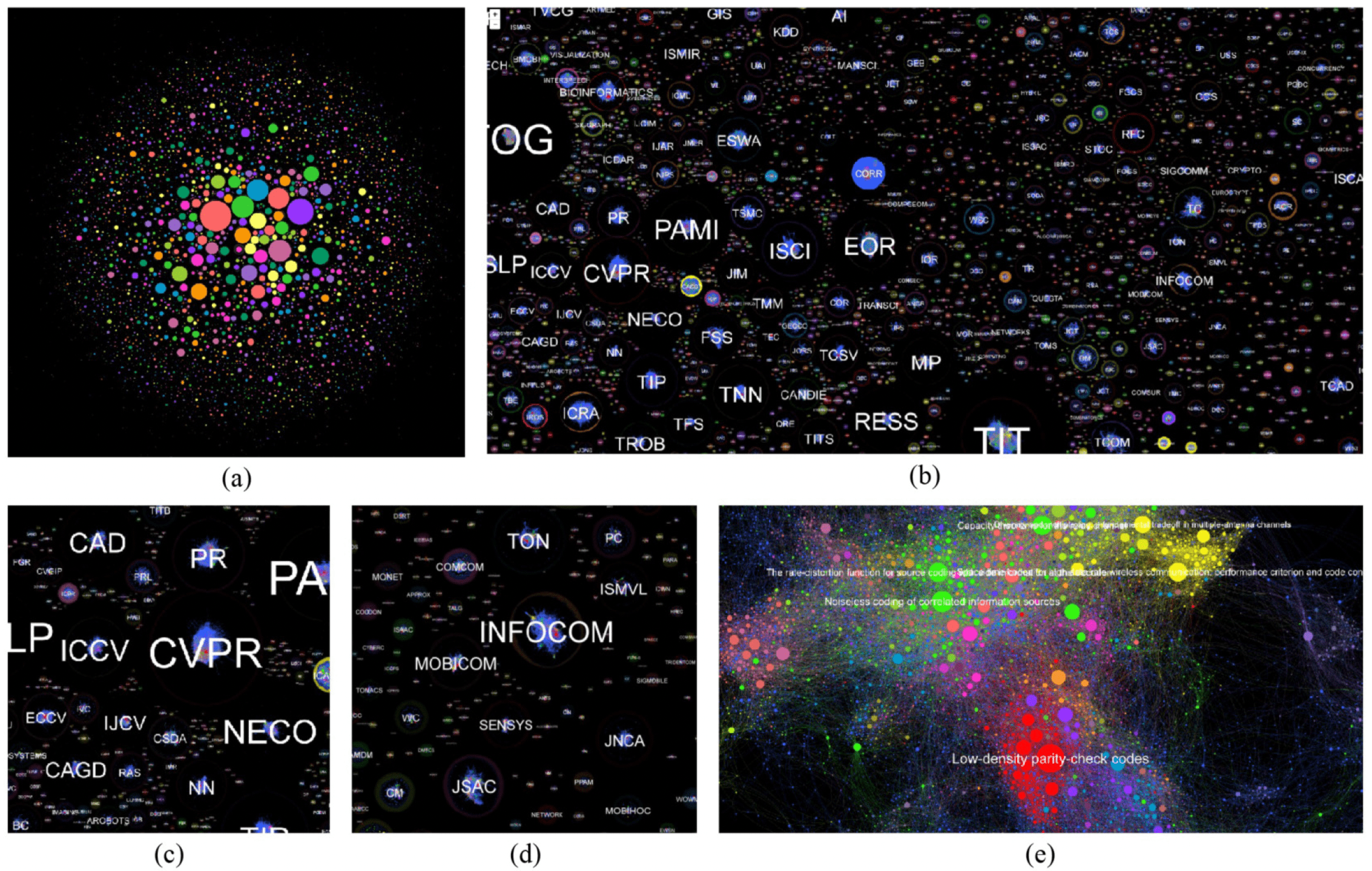}
    \caption{Visualizion of the entire DBLP papers network by VSAN}
    \label{fig:qili_f5}
\end{figure}

We have applied VSAN to visualize the entire DBLP papers network, which consists of more than 4.32M nodes and 36.03M edges (see Figure~\ref{fig:qili_f5}). In clustering the network, we did not employ a community discovery algorithm. Instead, we classified papers based on the conference or journal in which they were published. We also adapted formulas for attractive and repulsive forces between clusters. The de-overlap operation was integrated into the cluster layout process rather than being performed afterward. The final step of fine-tuning the overall graph was omitted.

The resulting galaxy-like structure of conferences and journals in the computer science field, visualized using VSAN, is displayed in Figure~\ref{fig:qili_f5}. Notably, top conferences in computer vision, such as CVPR, ICCV, and ECCV, can be clearly identified (Figure~\ref{fig:qili_f5}(c)). Similarly, the network in the computer networks field exhibits a convergence pattern (Figure~\ref{fig:qili_f5}(d)). Figure~\ref{fig:qili_f5}(e) reveals the clustering structure within TIT, a journal in the field of information theory, with different colors representing various topics. This structure displays a galaxy-like arrangement of papers within each topic, with larger nodes surrounded by smaller nodes of the same color.

\section{Quantification: Measure the amount of knowledge in papers}
\label{sec:quantification}

Researchers often need to discover high-quality and influential work from the vast amount of literature. Quantifying the amount of knowledge in academic papers can help them in this endeavor. In this part, we present a novel method, the Knowledge Quantification Index (KQI), which quantifies knowledge produced in academic literature, based on entropy theory. We also show how this method can provide new insights and perspectives on the scientific landscape.

\begin{figure}
	\centering
	\includegraphics[width=\linewidth]{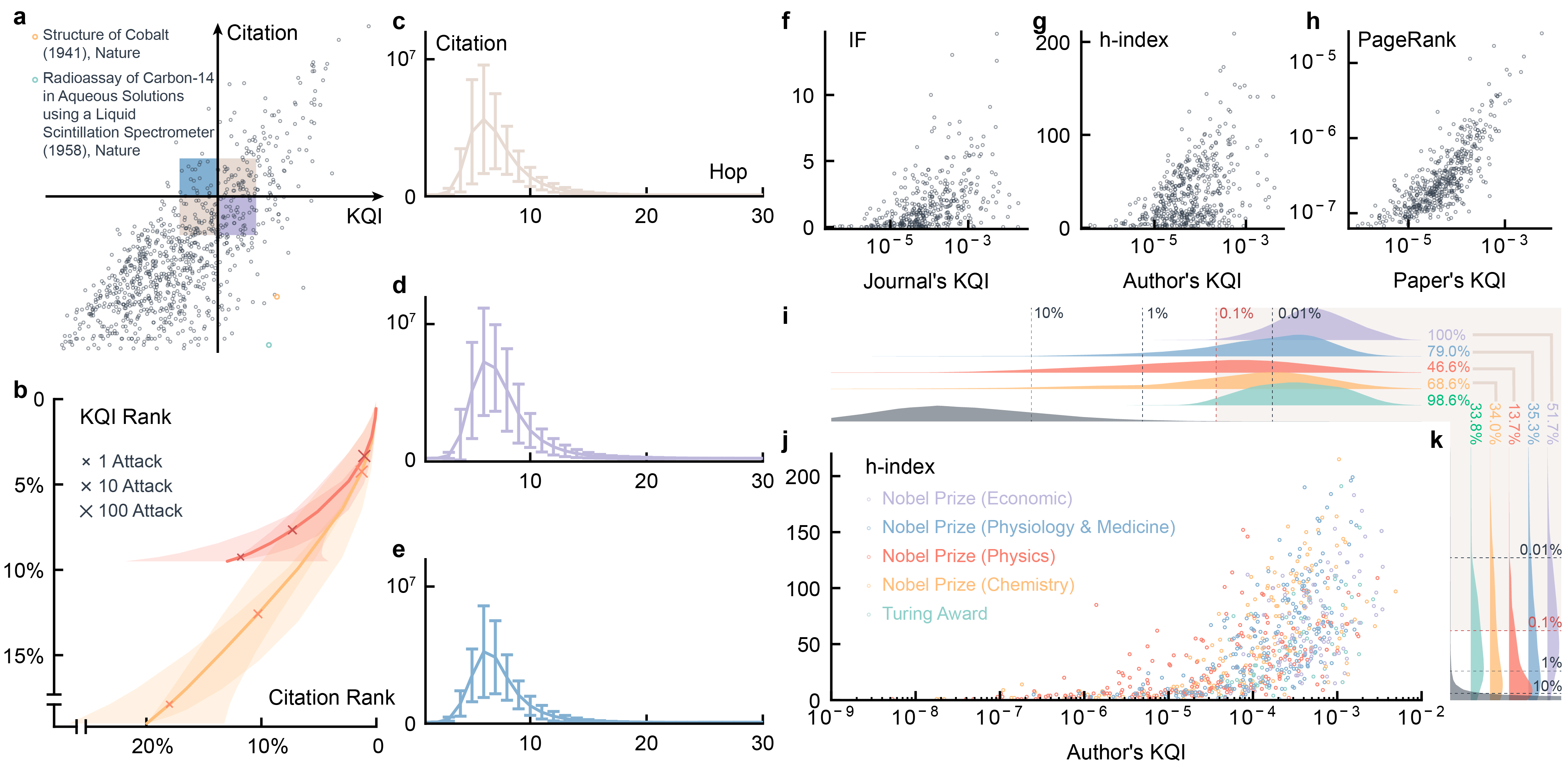}
	\caption[short]{Illustration of knowledge quantification.}
	\label{kqi}
\end{figure}

{\bf The Knowledge Quantification Index (KQI)} is a novel metric proposed to quantify knowledge produced in academic literature \cite{kqi}. 
It utilizes the hierarchical structure of citation networks to represent the extent of disorder difference caused by structure (order) and reflect the knowledge amount.
The formula of KQI can be written as follows: $$K = H^1 - H^T,$$ where the knowledge $K$ is always greater than 0, $H^1$ is Shannon entropy and $H^T$ is structural entropy \cite{Structural}.
The KQI is demonstrated to be effective in identifying valuable knowledge compared with traditional metrics (Fig. \ref{kqi}). 

We conduct analysis with KQI on 214 million articles published from 1800 to 2021 from various academic sources, including Nature, Science, Elsevier, and Springer, covering 292 fields in 19 disciplines. 
We created citation networks for these academic data and used the KQI metric to measure the knowledge in the network. 
The KQI can be used to reflect the knowledge attribute of the paper, i.e., the acceptability and dependability. 
Acceptability refers to whether knowledge is recognized, i.e., how much knowledge is inherited directly or indirectly from that knowledge. 
Dependability refers to whether the source of knowledge is equally or more recognized, i.e., how fully the parents can support the generation of the knowledge.
The higher the value of KQI, the stronger the two knowledge attributes are, i.e., high KQI papers are considered as justified truth, deriving from reliable parent knowledge and spawning numerous child knowledge. 

The KQI offers several advantages over existing measures. 
First, it takes into account the hierarchical structure of citation networks, which reflects the knowledge production process by idea flows from references to new papers.
Second, it is resistant to manipulation from self-citation and citation stacking and offers interpretability. 
Third, it can be used to identify valuable knowledge that is omitted by traditional metrics. 

Overall, the KQI offers a promising new approach to measuring knowledge in academic literature. 
Further research and application of the KQI can help improve the quality and impact of academic research and facilitate knowledge dissemination and discovery.

\paragraph{KQI of Turing Award winners} In this experimental dataset in computer science, there are 6,253,122 authors. By 2020, there are 74 Turing Award winners, and thirty percent of the top 50 authors according to KQI are Turing Award winners, while the remaining 70 authors are also highly influential and receive honors such as the IEEE John von Neumann Medal, MacArthur Fellows Program, or MacArthur Fellowship, Frederick W. Lanchester Prize, etc. We single out the top 10,000 authors (0.16\%) according to KQI (only first author concerned) and find 71 Turing Award winners (96\%) among them. The remaining three authors are Alan Perlis, James H. Wilkinson, and Kristen Nygaard. In the case of Alan Perlis, due to some objective factors, we do not have his representative works in this dataset. For James H. Wilkinson, his outstanding contribution in numerical analysis is classified into the field of mathematics by our dataset. Kristen Nygaard, who co-invented object-oriented programming and the Simula programming language with Ole-Johan Dahl, is listed as the second author. Kristen Nygaard falls behind in the rankings because we only consider the first author when dealing with the rankings.

\begin{figure}[h]
\centering
\includegraphics[width=\linewidth]{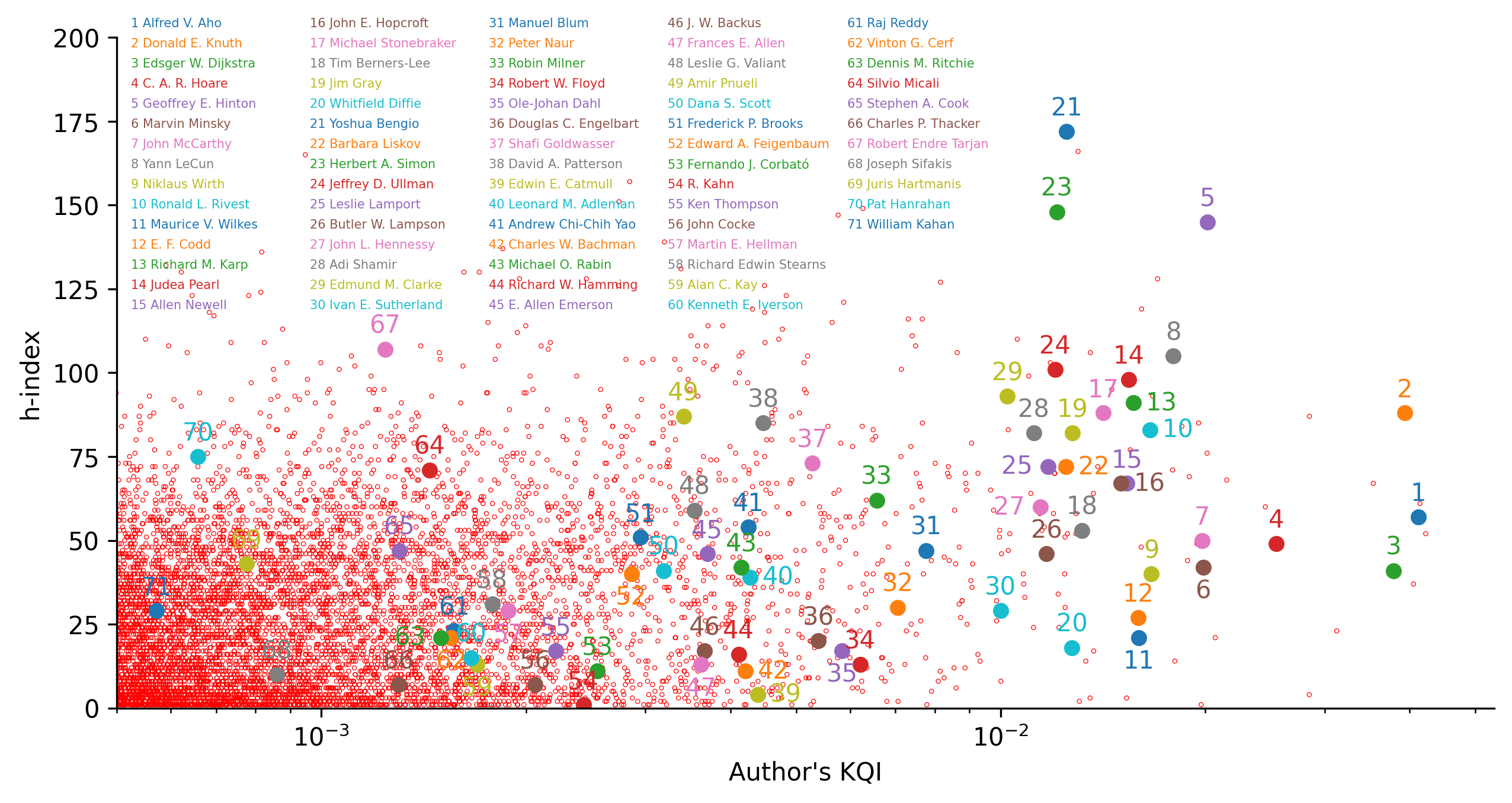}
\end{figure}

\paragraph{Countries Ranking by KQI} According to the country of papers’ first author, the top 20 countries are listed by aggregating KQI by country, with literature marked either.

\begin{figure}[h]
\centering
\includegraphics[width=0.8\linewidth]{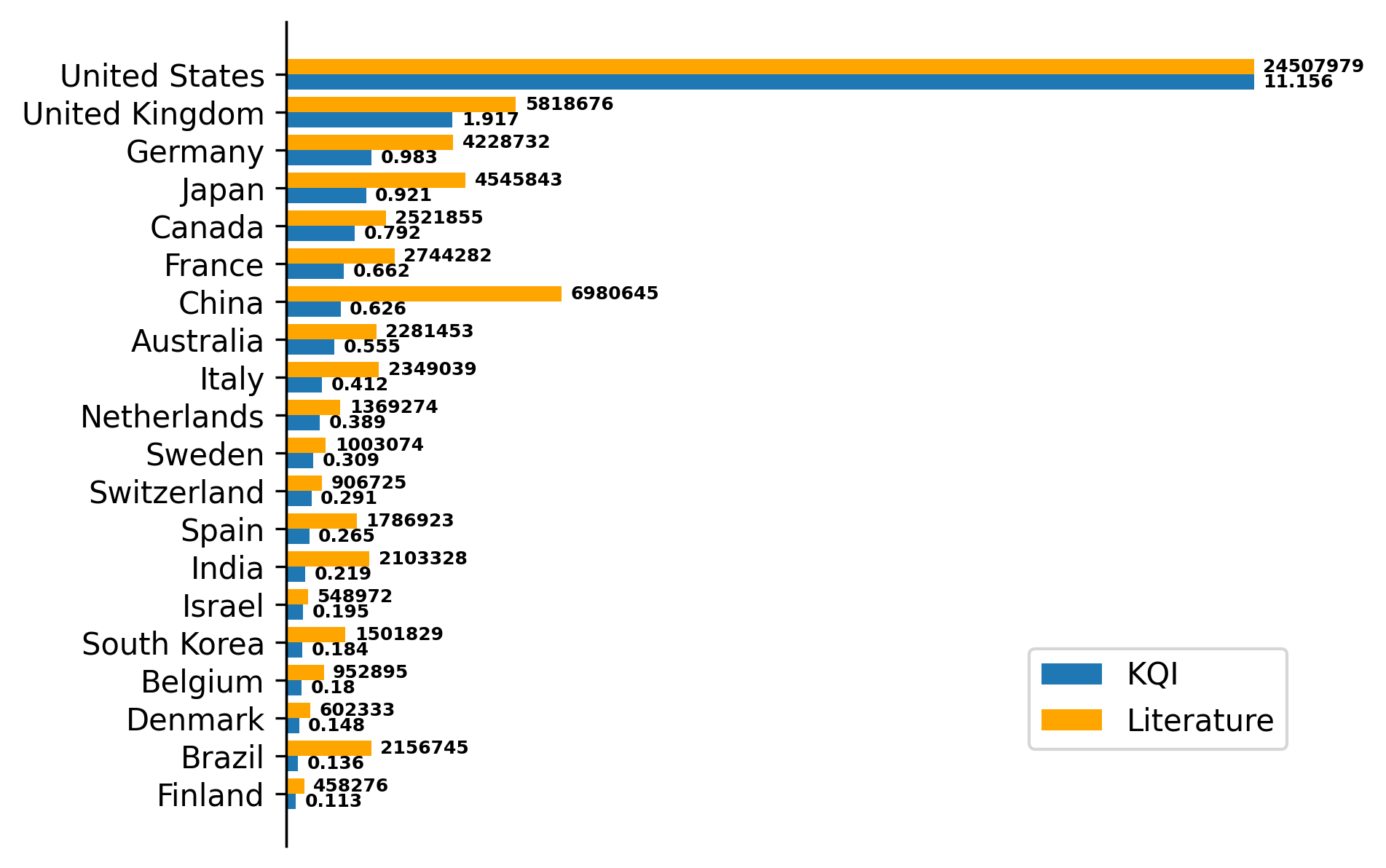}
\end{figure}

\paragraph{Affiliations Ranking by KQI} According to the affiliation of papers’ first author, the top 100 affiliations are listed by aggregating KQI by affiliations.

\begin{table}[h]
\centering
\caption{Affiliations Ranking by KQI.}
\resizebox{1.0\linewidth}{!}{%
\begin{tabular}{rp{6cm}rrp{6cm}r}
\hline
 No. &                    Name &     KQI ($\times 10^{-4}$) &  No. &                         Name &     KQI ($\times 10^{-4}$) \\
\hline
   1 &                               Harvard University &  4314.055 &   51 &                                              IBM &  644.219 \\
   2 &                              Stanford University &  3073.220 &   52 &                       Carnegie Mellon University &  623.605 \\
   3 &                    National Institutes of Health &  2375.006 &   53 &                                Purdue University &  606.947 \\
   4 &            Massachusetts Institute of Technology &  2207.661 &   54 &                                McGill University &  603.696 \\
   5 &               University of California, Berkeley &  2193.094 &   55 &                                Boston University &  594.303 \\
   6 &                           University of Michigan &  1845.644 &   56 &                 University of California, Irvine &  591.930 \\
   7 &                          University of Cambridge &  1822.393 &   57 &                          University of Edinburgh &  586.073 \\
   8 &                              Columbia University &  1798.909 &   58 &                         University of Manchester &  586.070 \\
   9 &            University of California, Los Angeles &  1782.025 &   59 &                               University of Utah &  580.203 \\
  10 &                         University of Washington &  1740.124 &   60 &                          University of Melbourne &  573.738 \\
  11 &                                  Yale University &  1589.931 &   61 &                                 Osaka University &  565.124 \\
  12 &                               Max Planck Society &  1563.456 &   62 &                            Karolinska Institutet &  552.533 \\
  13 &                            University of Chicago &  1499.324 &   63 &                             Texas A\&M University &  548.657 \\
  14 &                       University of Pennsylvania &  1496.673 &   64 &                          University of Rochester &  540.651 \\
  15 &                  University of Wisconsin-Madison &  1409.335 &   65 &  University of Texas Southwestern Medical Center &  537.137 \\
  16 &                               Cornell University &  1401.081 &   66 &                               University of Iowa &  537.066 \\
  17 &              University of California, San Diego &  1373.094 &   67 &                Spanish National Research Council &  526.036 \\
  18 &                          University of Minnesota &  1368.138 &   68 &                        Michigan State University &  523.051 \\
  19 &                         Johns Hopkins University &  1313.751 &   69 &                  Case Western Reserve University &  515.382 \\
  20 &                             University of Oxford &  1309.242 &   70 &                           University of Virginia &  512.322 \\
  21 &                             Princeton University &  1283.582 &   71 &                   University of Colorado Boulder &  506.716 \\
  22 &                              New York University &  1172.194 &   72 &                            Vanderbilt University &  506.472 \\
  23 &          University of California, San Francisco &  1165.678 &   73 &                         University of Copenhagen &  500.675 \\
  24 &                            University of Toronto &  1161.220 &   74 &                                  Lund University &  499.341 \\
  25 &                                  Duke University &  1138.047 &   75 &                     Brigham and Women's Hospital &  493.569 \\
  26 &     Centre national de la recherche scientifique &  1115.222 &   76 &                           Rockefeller University &  473.213 \\
  27 &       University of Illinois at Urbana–Champaign &  1078.021 &   77 &                             University of Sydney &  472.606 \\
  28 &                          Northwestern University &  1052.433 &   78 &                                 Brown University &  467.860 \\
  29 &                    University of Texas at Austin &  1034.175 &   79 &                         Arizona State University &  463.401 \\
  30 &               Washington University in St. Louis &   981.890 &   80 &                          University of Amsterdam &  460.407 \\
  31 &                        University College London &   971.396 &   81 &                            King's College London &  456.631 \\
  32 &                University of Southern California &   957.277 &   82 &          University of California, Santa Barbara &  455.662 \\
  33 &                              University of Tokyo &   937.604 &   83 &                   Hebrew University of Jerusalem &  455.032 \\
  34 &                      Chinese Academy of Sciences &   903.581 &   84 &           Ludwig Maximilian University of Munich &  448.234 \\
  35 &               California Institute of Technology &   902.551 &   85 &                               Utrecht University &  447.436 \\
  36 &      University of North Carolina at Chapel Hill &   882.683 &   86 &                              University of Paris &  446.511 \\
  37 &                         University of Pittsburgh &   853.370 &   87 &                                 Emory University &  443.105 \\
  38 &                    Pennsylvania State University &   833.428 &   88 &                   Katholieke Universiteit Leuven &  439.741 \\
  39 &                            Ohio State University &   810.434 &   89 &                            University of Bristol &  435.826 \\
  40 &                  University of California, Davis &   798.666 &   90 &                             University of Zurich &  434.222 \\
  41 &                               Rutgers University &   759.650 &   91 &                         University of Queensland &  434.211 \\
  42 &                            University of Florida &   743.100 &   92 &                   Australian National University &  424.273 \\
  43 &                   University of British Columbia &   728.339 &   93 &                       Baylor College of Medicine &  420.760 \\
  44 &                          Imperial College London &   716.584 &   94 &                            University of Alberta &  419.484 \\
  45 &  French Institute of Health and Medical Research &   712.008 &   95 &              University of Massachusetts Amherst &  402.536 \\
  46 &             University of Maryland, College Park &   701.468 &   96 &                               Uppsala University &  402.314 \\
  47 &                                      Mayo Clinic &   666.846 &   97 &                                Tohoku University &  399.097 \\
  48 &                                        Bell Labs &   664.646 &   98 &   National Institute of Standards and Technology &  393.813 \\
  49 &                                 Kyoto University &   662.686 &   99 &    University of Texas MD Anderson Cancer Center &  393.462 \\
  50 &                            University of Arizona &   646.425 &  100 &                              University of Miami &  390.988 \\
\hline

\end{tabular}}
\end{table}

\begin{figure}
    \centering\textbf{}
    \includegraphics[width=0.6\linewidth]{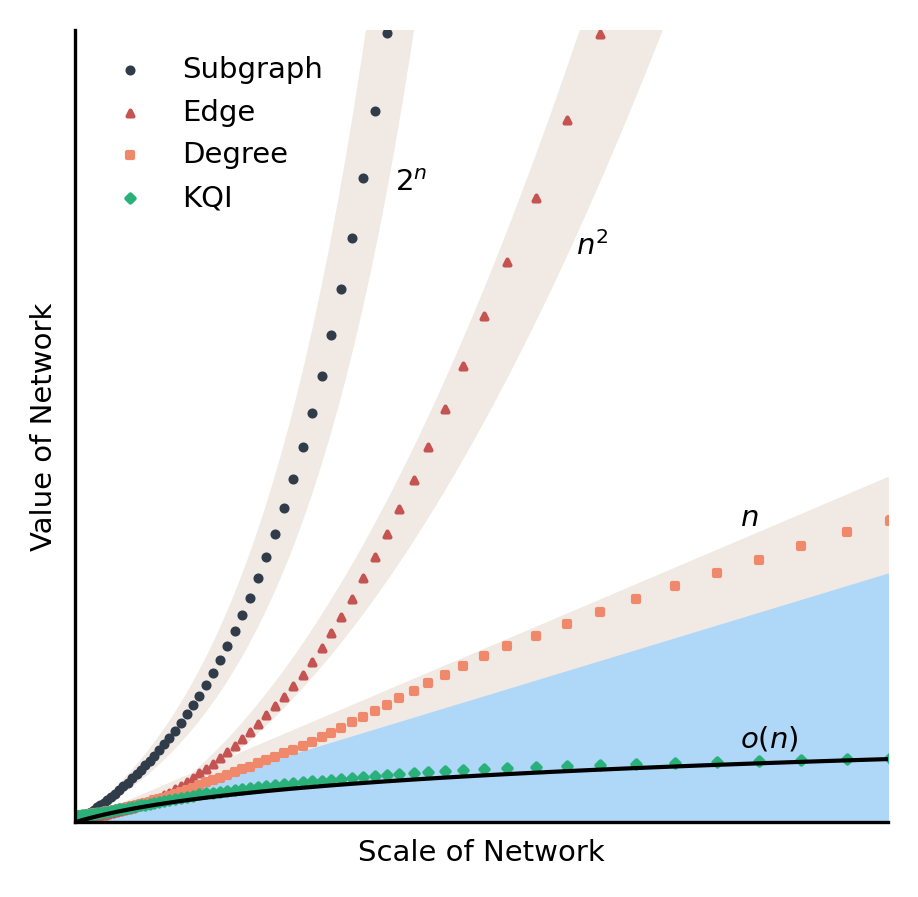}
    \caption{Sarnoff's law, Metcalfe's law, Reed's law, and Sublinear Knowledge law for measuring the value of different perspectives in citation networks.}
    \label{fig:network_value}
\end{figure}

\paragraph{Knowledge Growth Law} Sarnoff's Law\cite{SWANN2002417}, Metcalfe's Law\cite{metcalfe1976ethernet}, and Reed's Law\cite{reed2001law} were first proposed separately to represent network value in different contexts. We use the citation network constructed from the Deep-time Digital Earth (DDE)\cite{DDE} academic literature to identify the coexistence of three laws in citation networks. The edges emanating from a given node in the citation network represent the direct influence of that paper on other papers, according to Sarnoff's law, which suggests a linear relationship between knowledge diffusion and network size. The transmission of knowledge between papers is demonstrated by the diffusion value of the citation network. Metcalfe's Law is reflected in citation networks by the number of edges, which represents the total number of citations for all papers and is proportional to the square of the number of papers. Reed's law can be interpreted as the number of subgraphs in a graph, representing the possible set of literature that could be formed by combining papers, reflecting the diversity and possibility of creativity. Using KQI to calculate the amount of knowledge contained in each annual network snapshot of the citation network, we found that the knowledge contained in the network grows sublinearly as the network size increases, as shown in Figure~\ref{fig:network_value}. For a network with $n$ nodes, the amount of knowledge it contains is $o(n)$. The sublinear knowledge law illustrates the disparity between the growth of network size and the growth of the amount of knowledge contained in the network, meaning that a large amount of literature does not yield much knowledge.

\section{Analysis: Intelligent Academic Service}
\label{sec:analysis}

\subsection{IdeaReader: Understanding the Idea Flow of Scientific Publications}
Understanding the origin and impact of the idea of publications is essential for scientific research. However, the sheer volume of scientific publications presents a challenge in tracing the evolution of relevant literature. To address this issue, we introduce IdeaReader, a machine reading system designed to identify papers that are likely to inspire or be influenced by a target publication \cite{li2022ideareader}. IdeaReader also summarizes the ideas within these papers using natural language processing techniques. The algorithm pipeline for IdeaReader is outlined as follows (refer to Figure~\ref{fig:qili_f2}):

\begin{figure}
    \centering\textbf{}
    \includegraphics[width=1.0\linewidth]{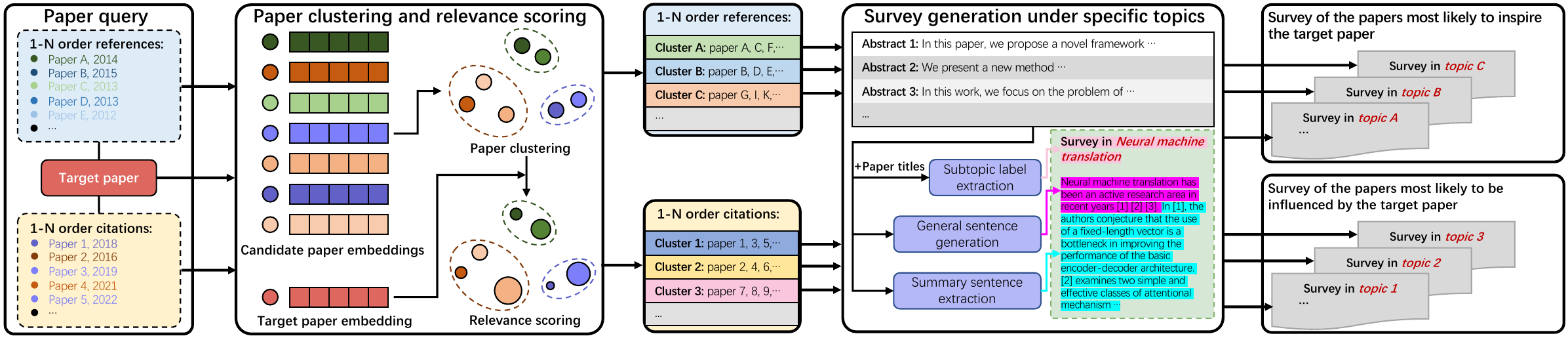}
    \caption{The pipeline of IdeaReader}
    \label{fig:qili_f2}
\end{figure}

\begin{figure}
    \centering
    \includegraphics[width=1.0\linewidth]{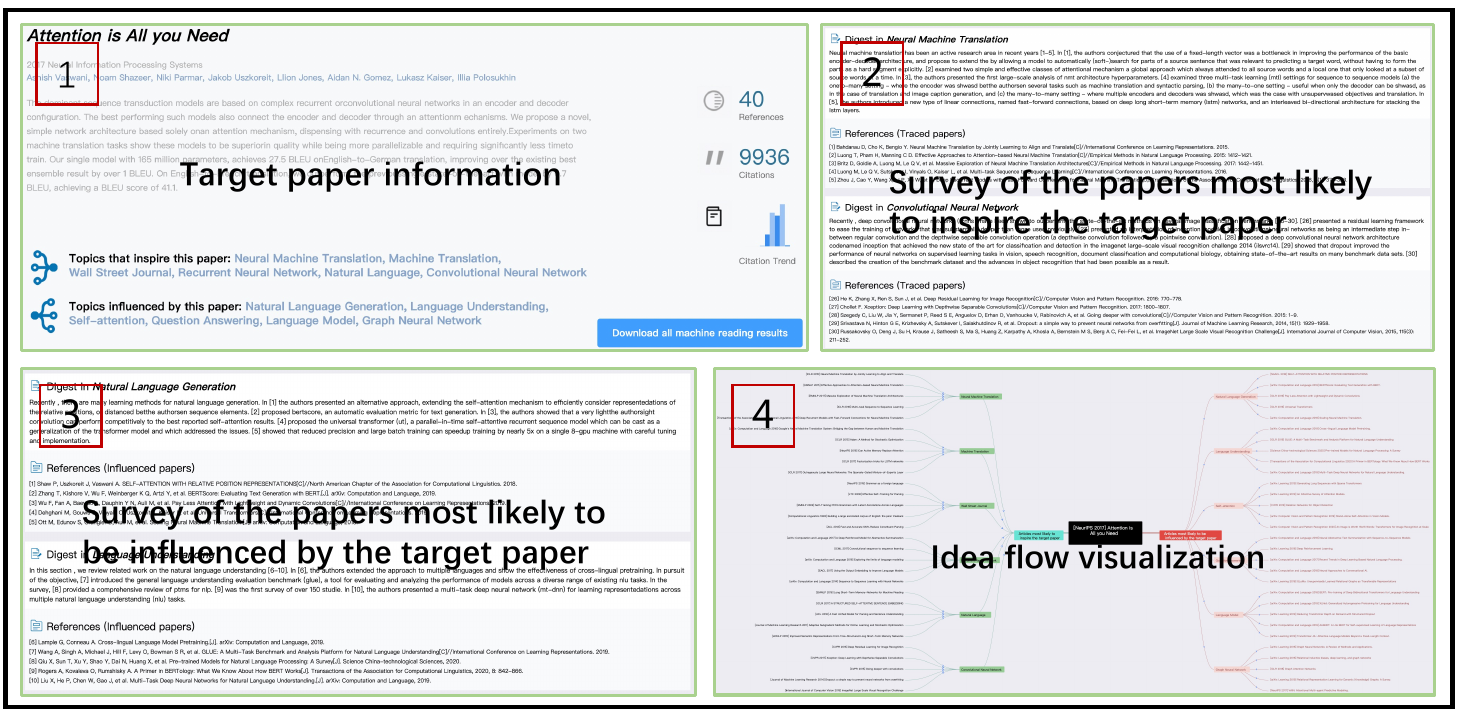}
    \caption{The front-end interface of IdeaReader}
    \label{fig:qili_f3}
\end{figure}

\paragraph{Paper query} Initially, our system queries the AceMap database \cite{tan2016acemap} to identify candidate reference and citation papers related to the target publication. For reference papers, IdeaReader selects first-order reference papers, and if their count is below 100, it proceeds to select second-order reference papers (papers cited by first-order reference papers). PageRank \cite{page1999pagerank} is employed to identify the top 100 papers among all first-order and higher-order reference papers. A similar rule is followed for selecting candidate citation papers.

\paragraph{Paper clustering and relevance scoring} Next, we utilize a method similar to MRT \cite{yin2021mrt} to cluster and rank candidate reference and citation papers separately. IdeaReader combines TF-IDF with Sentence-BERT \cite{reimers2019sentence} to encode paper abstracts, incorporating the resulting embeddings with ProNE \cite{zhang2019prone} based on the citation structure. Kernel k-means \cite{kulis2009semi} is then used to cluster candidate papers, treating each cluster as a topic associated with the target publication. To identify the papers most likely to inspire or be influenced by the target publication within a specific topic, IdeaReader calculates relevance scores using vector inner products and actual citation relationships. Finally, the system ranks papers within each cluster based on their relevance scores.

\paragraph{Survey generation under specific topics} Subsequently, IdeaReader automatically generates summaries of the top five papers with the highest relevance scores in each cluster. These surveys include a heading, a general sentence, and summary sentences from the selected papers. The system employs automatic text annotation \cite{mei2007automatic} to extract labels from the titles and abstracts of the selected papers, using them as headings for the corresponding topic surveys. The general sentence is generated from the abstracts of selected papers using BertSumABS \cite{liu2019text}, with fine-tuning based on the related work generation dataset from \cite{chen2021capturing} (2021). For summary sentences, IdeaReader employs SciBERT \cite{beltagy2019scibert} to train a binary classification model that extracts these sentences from the abstracts of input articles. The dataset for this purpose is constructed from the scientific abstract sentence classification datasets provided by \cite{cohan2019pretrained} (2019) and \cite{gonccalves2020deep} (2020), where sentences labeled as the 'OBJECTIVE' class are used as summary sentences. The generated text is aligned for readability in terms of subject and tense.

\paragraph{Front-end interface} Our front-end interface is shown in Figure~\ref{fig:qili_f3} and consists of four parts: (1) Target Paper Information Panel: This section provides metadata for the target paper, including references and citation statistics, as well as topics that inspire or are influenced by the target publication. (2) Survey of Papers Inspiring the Target Paper: This panel presents surveys on multiple topics, each within a single card, showcasing papers that are likely to inspire the target paper. (3) Survey of Papers Influenced by the Target Paper: Similar to the previous panel, this section features surveys on multiple topics, each in a single card, highlighting papers that are likely to be influenced by the target paper. (4) Idea Flow Visualization: known as the 'Tracing and Evolution Tree,' this visualization depicts the target paper as the root node, with the left branch showcasing topics and papers inspiring the target paper, and the right branch displaying topics and papers influenced by the target paper. The direction of idea flow is from left to right.

\subsection{DeepReport: Discovering Innovative Concepts in Scientific Literature}

\begin{figure}
    \centering
    \includegraphics[width=0.7\linewidth]{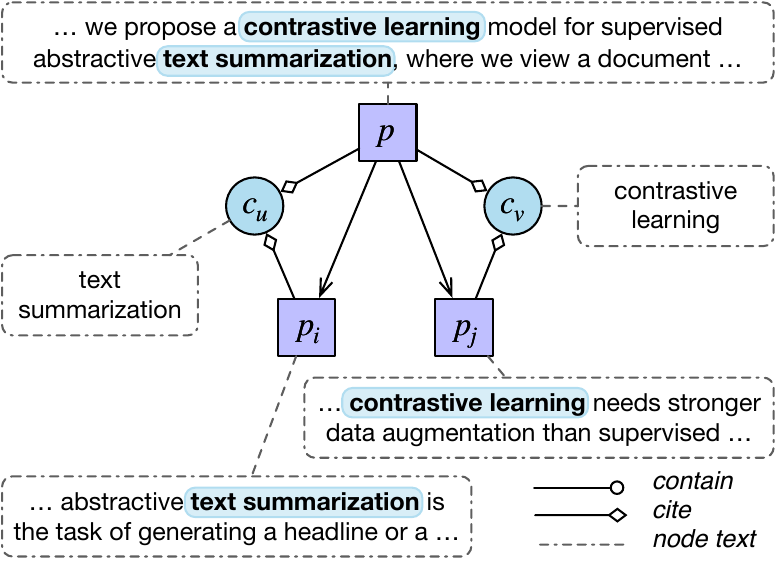}
    \caption{A quintuple with its text attributes. The dashed line and box represent the texts of paper or concept.}
    \label{fig:quintuple}
\end{figure}

Academic publications have observed human civilization's progress and advancement. In today's world, where a brilliant concept is the cornerstone of academic research, out-of-the-box, and interdisciplinary scientific study can garner more attention from science funders, industry, and the public. However, most researchers take a long time to present new ideas. The quantity of scholarly publications is growing at an exponential rate. Furthermore, scholars frequently focus on their specialized but confined domains, making it difficult to identify underlying relationships beyond their known areas. To this end, we introduce a novel research assistant system called DeepReport, seamlessly integrated into AceMap, to foster academic idea generation through concept co-occurrence analysis according to \cite{xu-etal-2023-exploring}.

Our approach hinges on the notion that the convergence of two concepts found within the pages of an academic paper can serve as a catalyst for the creation of new ideas. Hence we meticulously construct dynamic concept graphs based on the co-occurrence patterns observed across 20 distinct disciplines and topics. Subsequently, we employ a cutting-edge temporal link prediction method, leveraging masked language models, to unveil latent connections between diverse concepts.

To articulate these newfound associations, we harness pre-trained language models to craft descriptions of ideas using a groundbreaking data structure we term the "co-occurrence citation quintuple." as shown in Figure ~\ref{fig:quintuple}, where $p$ is called target paper, $p_i$ and $p_j$ are called reference papers. This innovative approach facilitates the articulation of novel linkages between concepts in a coherent and intuitive manner.

We meticulously evaluate the efficacy of our proposed system, employing a dual assessment approach that includes both automated metrics and human judgment. Our findings unequivocally affirm the promising potential of this system within the research landscape, as it significantly expedites the process of idea discovery for researchers.

\begin{tcolorbox}
\textbf{Knowledge Conclusion by DeepReport}\\

\textbf{Query}: AlphaFold.\\

\textbf{Answer}: AlphaFold is an artificial neural network that can be used to predict complex interactions between two or more heteromeric interfaces in real-time. \\

AlphaFold, the deep learning algorithm developed by DeepMind, recently released the three-dimensional models of the whole human proteome to the scientific community. \\

An AlphaFold model is trained specifically for multimeric inputs of known stoichiometry, which they call AlphaFold-Multimer, significantly increases accuracy of predicted multimeric interfaces over input-adapted single-chain AlphaFold while maintaining high intra-chain accuracy. \\

These researches describe the ability of AlphaFold to predict intermolecular interactions before and after single mutations in proteins with respect to their predicted changes in amino acid binding sites. The algorithms are highly accurate and have enabled unprecedented expansion of the structural coverage of the known protein-sequence space. \\

\end{tcolorbox}

\begin{figure}
    \centering
    \includegraphics[width=\linewidth]{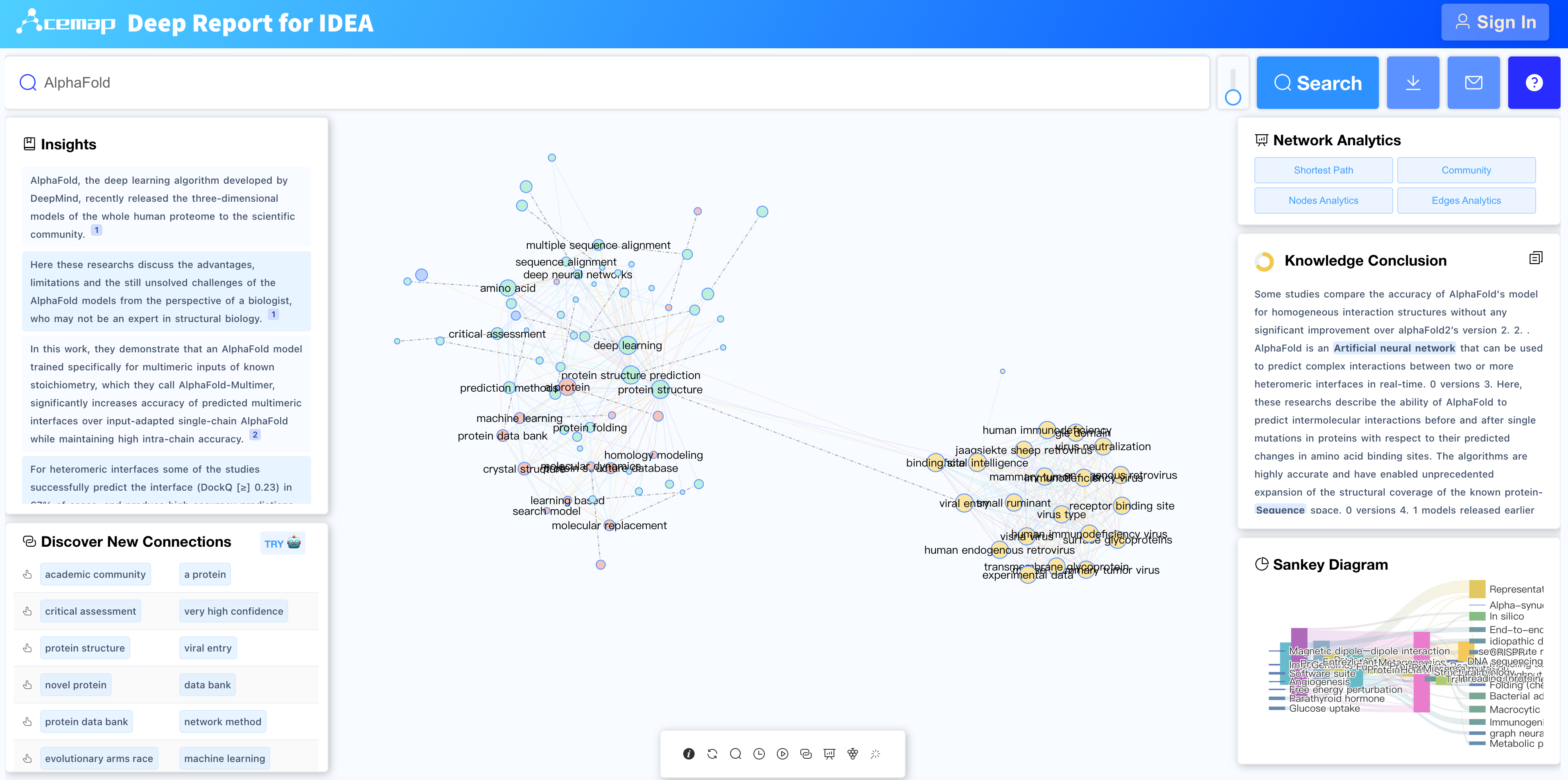}
    \caption{Result screenshot of DeepReport about the AlphaFold}
    \label{fig:enter-label}
\end{figure}

\section{Conclusion and Future Work}

The AceMap system represents a step forward in academic data mining and knowledge discovery. 
We have extracted large-scale entities, relationships, and figures, texts and tables from Open Access (OA) papers with the collaboration with geoscientists.
Expanding our efforts to academic articles from other fields brings a massive scale and a wealth of unexplored knowledge waiting to be discovered. In addition, the dynamic nature of the scientific literature and the continuous evolution of AI technologies provide numerous opportunities for future research and improvements. Here, we outline some key directions for further exploration and development of AceMap.
\par \textbf{Developing ``Large Knowledge Models'' for Understanding Papers.} In order to improve the understanding of scientific papers, future work should focus on the construction of a "large knowledge model" within AceMap. This model should integrate figures, tables, and data extracted from papers to provide a more comprehensive understanding and deeper insights into academic knowledge discovery.
\par \textbf{Enhanced Knowledge Mining in the Era of AI for Science.} The development of AI, especially large language models (LLMs), has opened up new possibilities for academic knowledge discovery in the era of AI for Science. Future work should harness the power of AI to improve the efficiency, accuracy, and depth of knowledge extraction and representation within AceMap, and to improve other various aspects of AceMap.
\par \textbf{Exploring the Evolution and Correlations of Academic Viewpoints.} Delving deeper into the scientific literature to uncover the evolution and correlations of academic viewpoints and insights is a promising avenue for future research and applications. To this end, effective measures for spatio-temporal analysis of the citation network and the user interaction network should be developed and verified within AceMap. 
\par \textbf{Ethical Considerations and Bias Mitigation.} As AceMap becomes increasingly influential in academic knowledge discovery, addressing ethical concerns and bias mitigation in knowledge representation and recommendations should be a key focus of future work. Ethical guidelines should be established to ensure transparency and accountability of the system's decision-making process.

\section*{Acknowledgement}

The authors extend their heartfelt gratitude to the individuals who made significant contributions throughout the development of AceMap. The authors also deeply appreciate the dedicated efforts of Weinan Zhang, Xiaohua Tian, Zhaowei Tan, Changfeng Liu, Yuning Mao, Yunqi Guo, Jiaming Shen, Ruijie Wang, Yuchen Yan, Jialu Wang, Yuting Jia, Ye Zhang, Junjie Ou, Hui Xu, Chong Zhang, and many others who have played instrumental roles in the advancement of AceMap.

\bibliographystyle{unsrt}
\bibliography{reference.bib}

\end{sloppypar}
\end{document}